\documentclass[useAMS,usegraphicx,usenatbib]{mn2e}
\usepackage{amssymb}

\title[Kinematics of powerful HH jets in Carina]{Kinematics of powerful jets from intermediate-mass protostars in the Carina nebula} 

\author[Megan Reiter and Nathan Smith]{Megan Reiter$^{1}$\thanks{E-mail: mreiter@as.arizona.edu} and Nathan Smith$^{1}$ \\
$^{1}$ Steward Observatory, University of Arizona, Tucson,
  AZ 85721, USA}

\begin{document}

\date{Accepted 2014 Mar ??. Received 2014 Mar ??; in original form 2014 March ??}

\pagerange{\pageref{firstpage}--\pageref{lastpage}} \pubyear{2014}

\maketitle

\label{firstpage}


\begin{abstract}
 We present measurements of proper motions and radial velocities of four powerful Herbig--Haro (HH) jets in the Carina nebula: HH~666, HH~901, HH~902, and HH~1066. 
Two epochs of \textit{Hubble Space Telescope} (\textit{HST}) imaging separated by a time baseline of $\sim 4.4$ years provide proper motions that allow us to measure the transverse velocities of the jets, while ground-based spectra sample their Doppler velocities. Together these yield full three-dimensional space velocities. 
Aside from HH~666, their identification as outflows was previously inferred only from morphology in images. 
Proper motions now show decisively that these objects are indeed jets, and confirm that the intermediate-mass protostars identified as the candidate driving sources for HH~666 and HH~1066 are indeed the origin of these outflows. 
The appearance of two new knots in the HH~1066 jet suggest recent ($\sim 35$ yr) changes in the accretion rate, underscoring the variable nature of accretion and outflow in the formation of intermediate-mass stars. 
In fact, kinematics and mass-ejection histories for all the jets suggest highly episodic mass loss, and point toward pronounced accretion fluctuations. Overall, we measure velocities similar to those found for low-mass protostars. However, the HH jets in Carina have higher densities and are more massive than their low-mass counterparts.  
Coarse estimates suggest that the heavy jets of intermediate-mass protostars can compete with or even exceed inject $\sim10$ or more times the cumulative momentum injection of lower-mass protostars. 
\end{abstract}



\section{Introduction}\label{s:intro}

Actively accreting protostars drive outflows that may aid in the removal of angular momentum and the dispersal of the protostellar envelope \citep{arc06}. 
Energy liberated from disc accretion powers the outflow, writing a record of the driving protostar's accretion history into the spatially resolved outflow structures. 
Knots in protostellar jets arise from changes in the ejection velocity, the mass-loss rate in the outflow, or both. 
Measurements of the spatially resolved jet density and outflow velocity can break this degeneracy, allowing jets that propagate mostly in the plane of the sky to be used as an indirect tracer of accretion variability \citep[e.g.][]{thi10}.

Inferring the accretion history from the density structure in the jet depends critically on the ability to measure all of the mass in the jet and its kinematics. 
Jets are typically traced via millimeter emission of entrained molecules \citep[e.g.][]{beu08,fue01}, shocked H$_2$ emission \citep[e.g.][]{sta02,zha13}, or shock-excited atomic emission lines from Herbig--Haro (HH) objects \citep[e.g.][]{wan09}. 
None of these provide a complete census of the mass in the jet. 
A collimated jet may power the wider-angle outflows seen at longer wavelengths \citep[see e.g.][]{har94}, but this largely neutral jet will remain invisible behind large columns of gas and dust in typical star forming regions. 
For the special case of jets driven into an H~{\sc ii} region created by nearby massive stars, however, external UV irradiation illuminates even unshocked material, allowing for a more complete census of the mass loss. 
Emission from these irradiated HH jets is interpreted using standard diagnostics of photoionized gas, rather than models of non-linear and time-dependent shocks \citep{bal01}, providing an unusual opportunity to measure the physical properties of the outflow directly. 
This is especially valuable in the inner jet, before deceleration by shocks alters the density and velocity structure. 

Detailed jet mass-loss histories of a population of intermediate-mass protostars can provide a time-resolved view of the mass assembly of stars in the poorly understood intermediate-mass range ($\sim 2-8$ M$_{\odot}$) where changes in the physics of formation between low- and high-mass stars -- if any -- would begin to manifest. 
The Carina nebula hosts the largest known population of more massive irradiated jets \citep{smi10}, which are likely driven by intermediate-mass protostars \citep{ohl12,rei13}. 
Estimates of the time-averaged mass-loss rate of the HH jets in Carina have been derived from the H$\alpha$ emission measure using narrowband \textit{Hubble Space Telescope (HST)} images 
and assuming a velocity of $200$ km s$^{-1}$ \citep{smi10}. 
Crucially, this method only traces mass-loss in the photoionized part of the jet, and thus provides a lower limit on the mass-loss rate. 

Recent narrowband near-IR imaging revealed bright [Fe~{\sc ii}] emission in all four of these HH jets \citep{rei13}. 
In the harsh environment of the Carina nebula, where many nearby O-type stars bathe these jets in UV radiation, we would expect Fe$^+$ to be rapidly ionized to Fe$^{++}$. 
The fact that [Fe~{\sc ii}] is bright in these jets indicates that this is the not the case, and therefore the [Fe~{\sc ii}] emitting gas must be self-shielded. 
This self-shielded gas is not traced by H$\alpha$ and it must be high density in order to protect neutral material in the jet. 
Accounting for this, revised estimates of the mass-loss rate are generally an order of magnitude higher than the value derived from the H$\alpha$ emission measure \citep{rei13}. 

Without kinematic measurements, one must assume a velocity to estimate the mass-loss rate, adding substantially to the uncertainty. 
This is especially true if more massive protostars drive faster jets \citep[e.g.][]{cor97}. 
\textit{HST} imaging over multiple epochs now makes it possible to measure the apparent motion on the sky of fast jet knots, and when combined with radial velocities from spectroscopy, allows us to measure the tilt angle from the plane of the sky and the three-dimensional space velocity. 

In this paper, we present three dimensional kinematics of four spectacular HH jets in the Carina nebula -- HH~666, HH~901, HH~902, and HH~1066. 
Using archival H$\alpha$ images from the Wide Field Camera 3 (WFC3-UVIS) on board \textit{HST} together with Advanced Camera for Surveys (ACS) H$\alpha$ images from \citet{smi10} provides a time baseline of $\sim 4.4$ years, sufficient to measure velocities $\gtrsim 50$ km s$^{-1}$ for faint nebular condensations at the distance of the Carina nebula \citep[2.3 kpc,][]{smi06b}. 
The velocities measured here provide an essential element for determining the spatially-resolved mass-loss rates and momenta of these outflows.


\section{Observations}\label{s:obs}

\begin{table*}
\caption{Observations}
\centering
\begin{tabular}{llllllll}
\hline\hline
Target & $\alpha_{\mathrm{J2000}}$ & $\delta_{\mathrm{J2000}}$ & 
Instrument/ & Filter & Date & Int. \\
 &  &  & Telescope &  &  & time (s) \\
\hline
HH~666 &  10:43:51.3  &  -59:55:21  &  ACS/\textit{HST}  &  F658N  &  2005 Mar 30  &  1000  \\ 
HH~666 &  10:43:51.3  &  -59:55:21  &  WFC3/\textit{HST}  &  F656N  &  2009 Jul 24-29  &  7920  \\ 
HH~666 &  10:43:51.3  &  -59:55:21  &  EMMI/\textit{NTT}  &  H$\alpha$, [S~II]  &  2003 Mar 11  &  2400  \\ 
\hline
HH~901  &  10:44:03.5  &  -59:31:02  &  ACS/\textit{HST}  &  F658N  &  2005 Jul 17  &  1000  \\
HH~901  &  10:44:03.5  &  -59:31:02  &  WFC3/\textit{HST}  &  F657N  &  2010 Feb 1-2  &  1980  \\
HH~901  &  10:44:03.5  &  -59:31:02  &  EMMI/\textit{NTT}  &  H$\alpha$, [S~II]  &  2003 Mar 10  &  1800  \\
\hline
HH~902  &  10:44:01.7  &  -59:30:32  &  ACS/\textit{HST}  &  F658N  &  2005 Jul 17  &  1000  \\
HH~902  &  10:44:01.7  &  -59:30:32  &  WFC3/\textit{HST}  &  F657N  &  2010 Feb 1-2  &  1980  \\
HH~902  &  10:44:01.7  &  -59:30:32  &  EMMI/\textit{NTT}  &  H$\alpha$, [S~II]  &  2003 Mar 10  &  1800  \\
\hline
HH~1066 &  10:44:05.4  &  -59:29:40  &  ACS/\textit{HST}  &  F658N  &  2005 Jul 17  &  1000  \\
HH~1066 &  10:44:05.4  &  -59:29:40  &  WFC3/\textit{HST}  &  F657N  &  2010 Feb 1-2  &  1980  \\
HH~1066 &  10:44:05.4  &  -59:29:40  &  FIRE/\textit{Magellan}  &  [Fe~{\sc ii}] 1.64\micron\  &  2012 Mar 07  &  2280  \\
\hline
\end{tabular}
\label{t:obs}
\end{table*}

\subsection{HST H$\alpha$ Images}
Table~\ref{t:obs} lists the details of the HST images and optical and near-IR spectroscopy used in this study. 
First-epoch images of HH~666 were obtained with ACS onboard \textit{HST} on 30 March 2005 \citep{smi10} using the F658N filter. 
Second epoch H$\alpha$ images of HH~666 were obtained with WFC3-UVIS using the F656N filter on 24 July 2009 as part of the Early Release Observations after SM4, providing a $4.32$ year baseline between epochs. 
Using slightly different filters for the two epochs could influence the proper motions if H$\alpha$ and [N~II] do not have the same spatial distribution. 
The F658N filter used with ACS contains both H$\alpha$ and [N~II] while the F656N filter used with WFC3-UVIS isolates H$\alpha$. 
H$\alpha$ and [N~II] may be offset from each other along the flow axis in purely shock-excited gas (as H$\alpha$ and [S~II] appear to be in HH~110, \citealt{nor96}). 
In the heavily irradiated environment of the Carina nebula, the jets are photoionized, and would be more likely to have stratified emission perpendicular to the direction of propagation \citep[see, e.g. HH~901,][]{rei13}. 
However, the ionization potentials of H$\alpha$ and [N~II] are similar (13.6 eV and 14.5 eV, respectively), which will diminish the spatial offset between the lines. 
Residuals on the order of the noise in a subtraction image of individual, aligned jet features suggests that using observations taken in two different filters does not significantly bias our proper motions. 

HH~901, HH~902, and HH~1066 are all within the Trumpler 14 mosaic imaged with ACS on 17 July 2005. 
These three jets were observed in H$\alpha$ a second time with WFC3-UVIS on 1-2 February 2010 providing a time baseline of $4.55$ years. 
Both epochs of the HH~901 mosaic were observed with filters that contain H$\alpha$ and [N~II]. 

Starting with reduced (including distortion correction), calibrated, and drizzled mosaic images from the Hubble Legacy Archive\footnote{http://hla.stsci.edu/hlaview.html}, WFC3-UVIS images were registered to the ACS images in two steps. 
The WFC3-UVIS H$\alpha$ image was resampled to the ACS pixel scale using bilinear interpolation, then shifted relative to the ACS image to minimise the residuals in a subtraction image. 
RMS deviations of the stellar centroids of the optimally aligned images \citep[determined using {\sc Source Extractor},][]{ber96} are 0.2 pixels ($\sim 10$ mas), corresponding to a velocity uncertainty of $\sim 25$ km s$^{-1}$ for a time baseline of $\sim 4.4$ years. 
We adopt the distance to $\eta$ Carinae of 2.3 kpc \citep[$\pm 50$ pc, measured from the expansion of the Homunculus nebula;][]{smi06b} for all four jets (see also \citealt{smi07}). 

We select bright knots that do not change significantly in brightness and morphology in order to measure proper motions in the jet.
After subtracting a median-filtered image (with box size optimised for each jet feature) as a model of the background emission, we use the modified cross-correlation technique described by \citet[][see also \citealt{mor01} and \citealt{har01}]{cur96} to measure the transverse velocities of these nebulous jet features. 
Briefly, in this method a cross-correlation array is generated by shifting a small box containing a jet feature relative to a reference image. 
For each offset, the total of the square of the difference between the two images is computed. 
The minimum value of this array corresponds to the best match of the two images, and thus the pixel offset corresponding to the proper motion of the object. 

Systematic uncertainties affect all proper motion measurements and are of particular concern when using data obtained with two different instruments especially those located in different places on the focal plane that require different distortion corrections, as is the case with ACS and WFC3. 
A larger source of uncertainty comes from the bright and highly spatially-dependent nebulosity near the jets in Carina. 
This significantly reduces the contrast between diffuse jet features and the background, leading to much lower signal-to-noise than obtained for bright, nearby HH jets \citep[e.g.][]{har01}. 
The typical uncertainty of our proper motion measurements is $\sim 20-25$ mas, constrained by comparing the subtraction residuals with the background. 
This translates to velocity uncertainties ranging from $\sim 40-80$ km s$^{-1}$ (see Table~3) depending on the contrast between the jet feature and the background. Our proper motion measurements are somewhat more uncertain than what \citet{yus05} found applying a similar technique to measure proper motions of HH~399 using signal-to-noise limited images.

\subsection{Spectroscopy}
H$\alpha$ and [S~II] spectra of HH~902 and HH~901 were obtained with the ESO Multi-Mode Instrument (EMMI) on the NTT on 10 March 2003. 
High resolution, long-slit spectra of HH~902 were obtained in single-order, long-slit mode using two different narrowband order-sorting filters to isolate H$\alpha$ + [N~II] or [S~II] $\lambda\lambda$6717, 6731. 
The total slit length was $\sim 6$\arcmin. 
HH~901 was observed in echelle mode with the cross-dispersing grism allowing H$\alpha$ and [S~II] to be observed simultaneously with a slit length of 25\arcsec. 
Both jets were observed with a 1\arcsec\ slit width. 
The wavelength solution was computed from the nebular [N~II] lines (for the H$\alpha$ spectrum) or from the [S~II] lines themselves. 
HH~666 was observed with the same setup as HH~902, and the resulting spectra were published by \citet{smi04}. 

No optical spectra were obtained for HH~1066. 
Since HH~1066 has extremely bright near-IR [Fe~{\sc ii}] emission \citep{rei13}, we observed HH~1066 with the Folded-Port InfraRed Echellette (FIRE) near-IR spectrograph \citep{sim08,sim10,sim13} on the Magellan Baade 6.5~m telescope on 07 March 2012. 
FIRE's $0.8 - 2.5$ \micron\ wavelength coverage includes multiple emission lines from the jet, including the [Fe~{\sc ii}] 1.644 \micron\ line which we use in this analysis. 
The 7\arcsec\ slit was aligned along the jet axis, centred on the bright [Fe~{\sc ii}] emission found by \citet{rei13}.  
We used a 0$\farcs$75 slit width to accommodate variable seeing conditions over the course of the night. 
To account for sky emission, we employed a nodding strategy, pointing on- and off-source in an ABBA sequence.  
The internal ThAr lamp was used for wavelength calibration and the data were reduced using the Firehose IDL pipeline.

\section{Results}\label{s:results}
\begin{figure*}
\centering
\includegraphics[trim=50mm 10mm 25mm 0mm,angle=-90,scale=0.575]{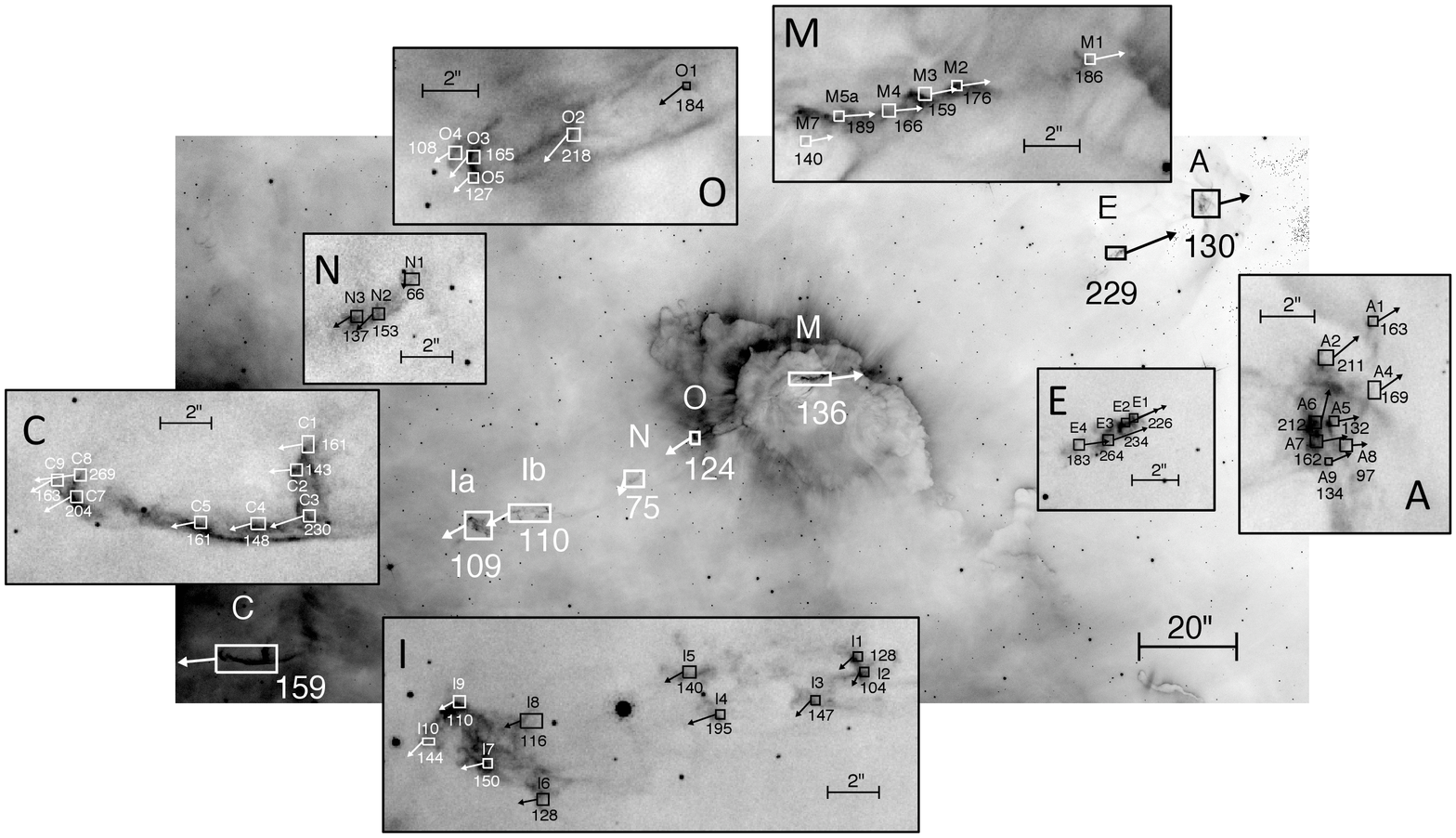} 
\caption{Boxes and velocity vectors representing the proper motions measured in HH~666 using HST images taken 4.32 years apart are superposed on the new WFC3-UVIS H$\alpha$ image. Features HH~666~A, E, M, O, N, I, and C, identified by \citet{smi04}, are labelled (HH~666~D is outside the area imaged with WFC3) along with the measured transverse velocity in km s$^{-1}$. 
Inset images show a zoomed view of the individual features HH~666~A, E, M, O, N, I, and C with boxes and velocity vectors showing the smaller knots within the larger features also measured.}\label{fig:hh666_pm} 
\end{figure*}
\begin{figure*}
\centering
\includegraphics[angle=0,scale=0.65]{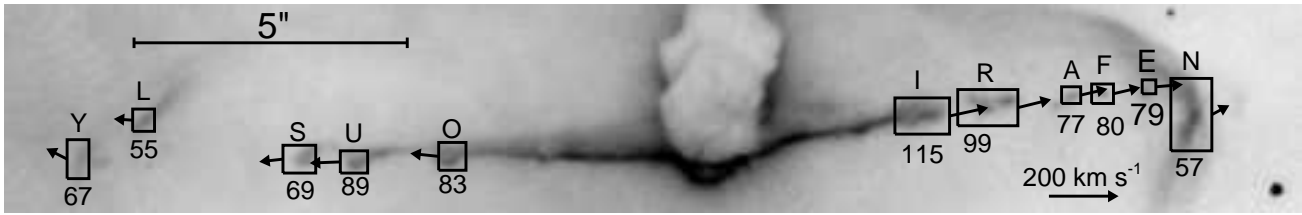}
\caption{WFC3-UVIS H$\alpha$ image of HH~901 with boxes indicating features used to compute proper motions and velocity vectors indicating the magnitude and direction of the measured motion.}\label{fig:hh901_pm} 
\end{figure*}
\begin{figure*}
\centering
\includegraphics[angle=0,scale=0.55]{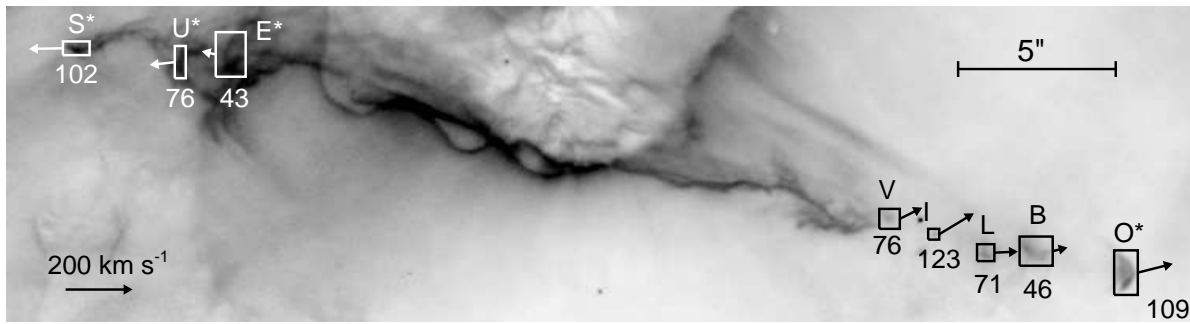} 
\caption{WFC3-UVIS image of HH~902 with boxes indicating the features used to measure proper motions. H$\alpha$ knots used to measure proper motions in HH~902 that overlap with the IR-bright features identified by \citet{rei13} are denoted with an $^*$. New feature labels were chosen within the constraints imposed by existing feature names. }\label{fig:hh902_pm} 
\end{figure*}
\begin{figure}
\centering
$\begin{array}{c}
\includegraphics[angle=0,scale=0.275]{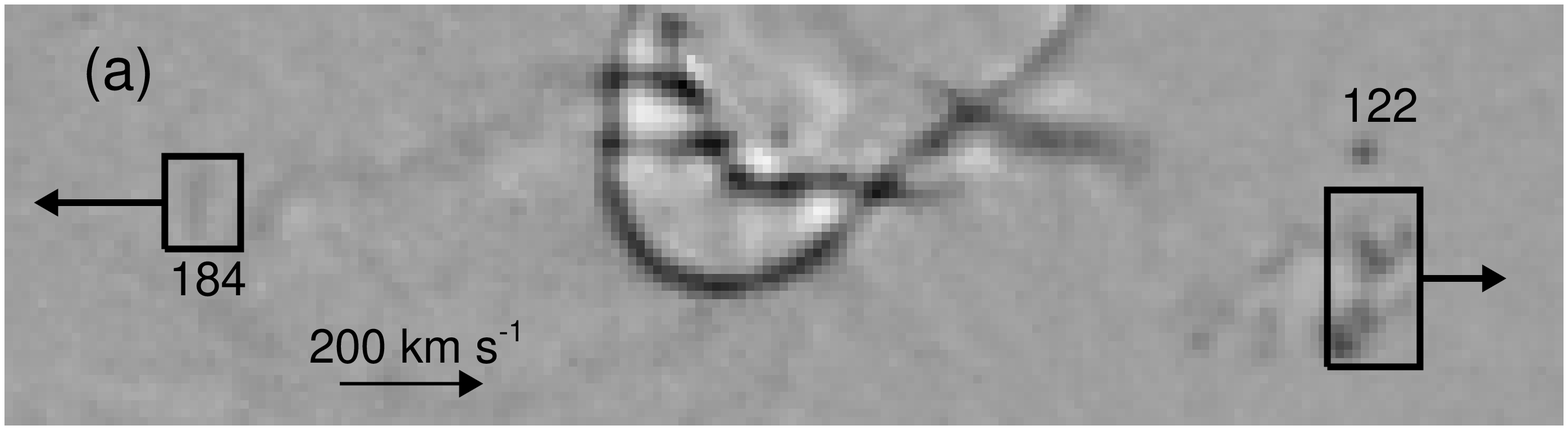} \\
\includegraphics[angle=0,scale=0.275]{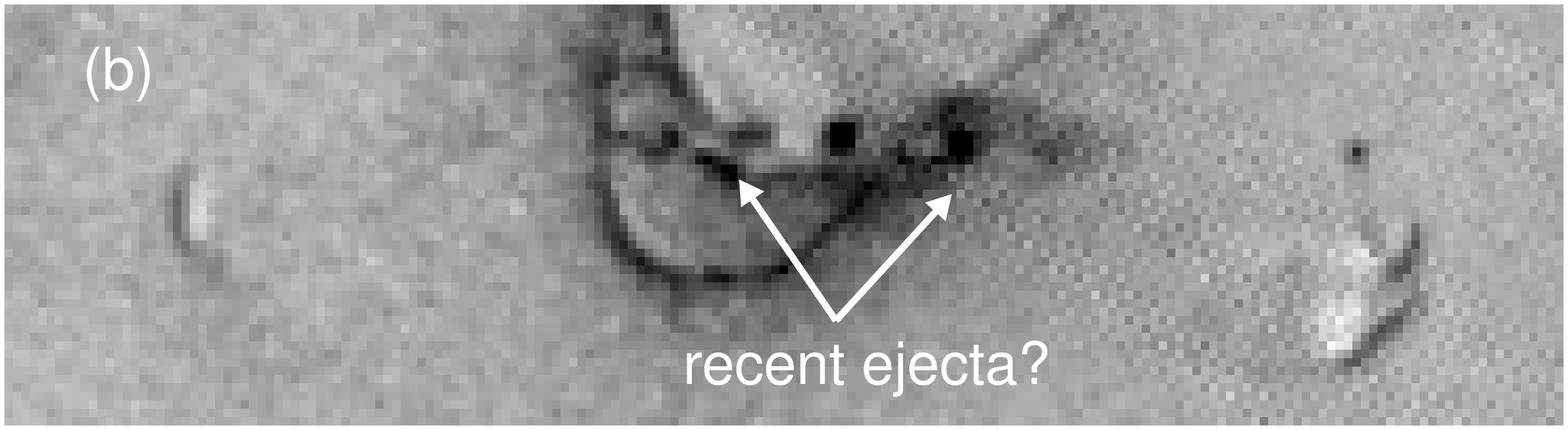} \\
\includegraphics[angle=0,scale=0.275]{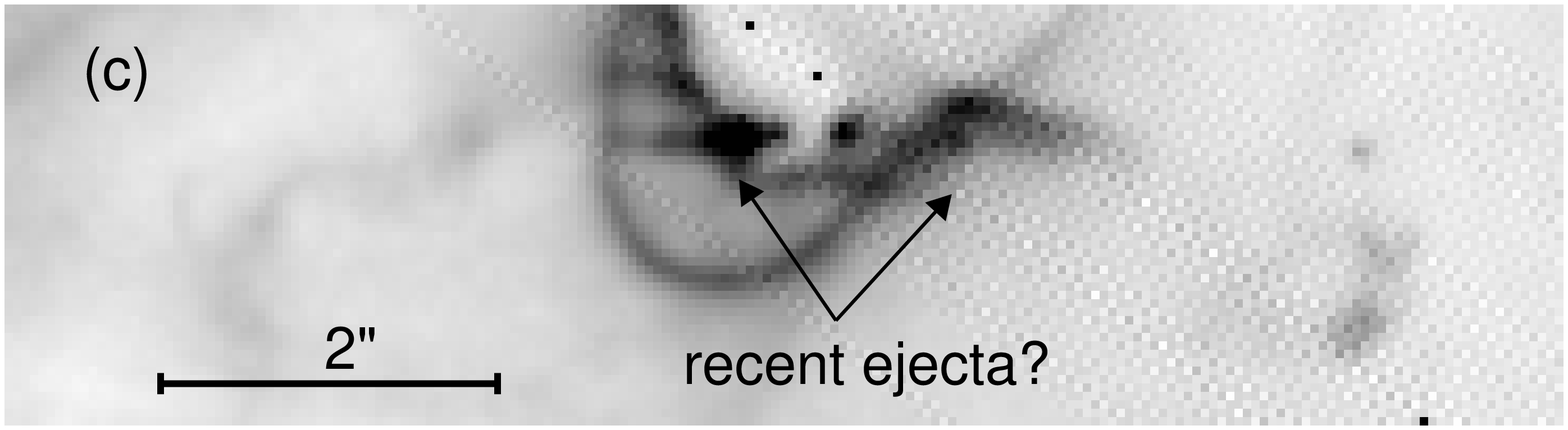} \\
\includegraphics[angle=0,scale=0.305]{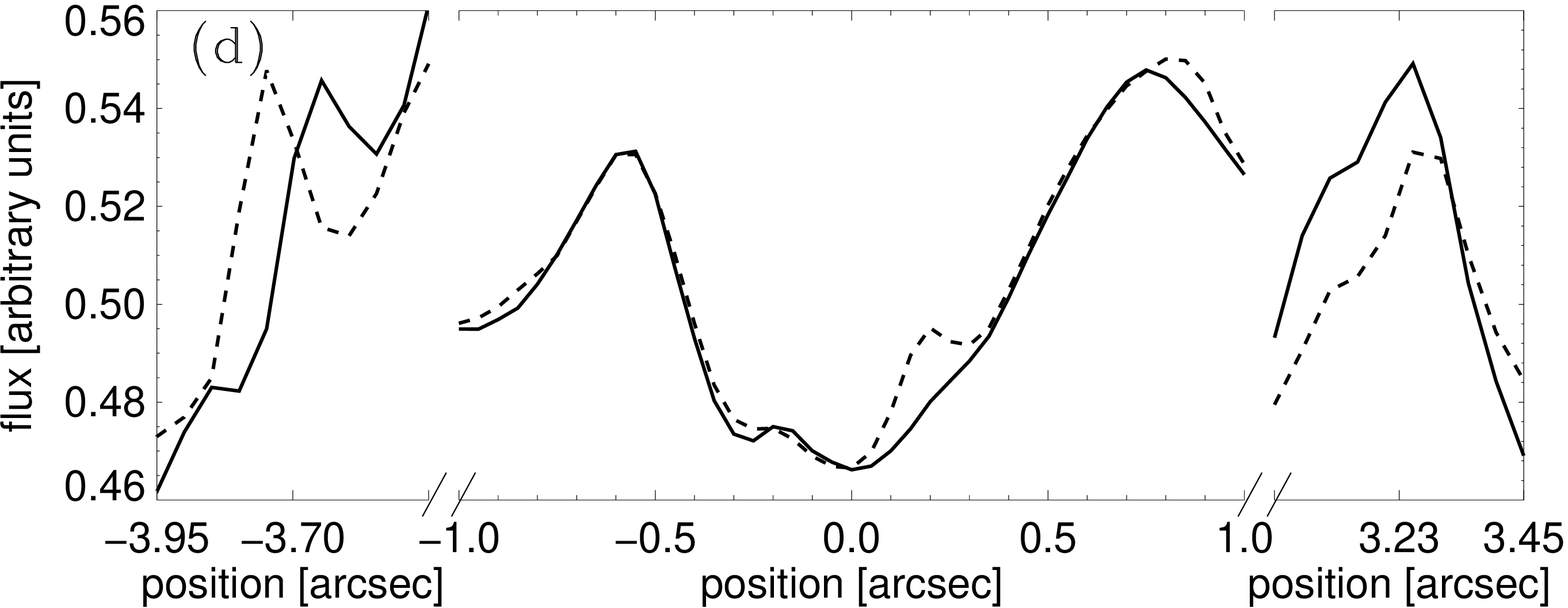} \\
\end{array}$
\caption{Proper motions of blobs in HH~1066 (a) shown with boxes and velocity vectors on the new WFC3-UVIS H$\alpha$ image, (b) in a difference image between the two epochs, (c) as bright knots in the WFC3-UVIS [S~II] image. Panels (d) show intensity tracings of the recent ejecta as well as the two bow shocks. }\label{fig:hh1066} 
\end{figure}

Combining transverse velocities from proper motions with radial velocities from spectra, we calculate the three-dimensional space velocities (total velocities) and orientations of HH~666, HH~901, HH~902, and HH~1066 in the Carina nebula. 
Derived jet kinematics are listed in Table~3. 
We calculate transverse velocities from the adopted distance to the Carina nebula (2.3 kpc) and translate these to representative dynamical ages assuming ballistic motion of the jet knots over this short time baseline. 
The jet features used to measure proper motions (see Figures~\ref{fig:hh666_pm}, \ref{fig:hh901_pm}, \ref{fig:hh902_pm}, and \ref{fig:hh1066}) show little change in morphology between the two epochs, suggesting a lack of significant turbulent motions in the jets that would compromise the measurements. 
For each jet, we define the tilt angle, $\alpha$, of the flow axis away from the plane of the sky as $\alpha = \mathrm{tan}^{-1} (v_r / v_T)$. 
All four of the jets have total velocities in excess of 100 km s$^{-1}$. 
In the following paragraphs, we briefly describe notable features of the individual jets.

\textit{HH~666:} 
Optical spectroscopy of HH~666 reveals a sawtooth velocity pattern with Doppler velocities up to $\pm 250$ km s$^{-1}$ \citep{smi04}. 
Transverse velocities measured from proper motions of the HH~666 blobs (see Figure~\ref{fig:hh666_pm}) are $75-229$ km s$^{-1}$, with velocities from the inner jet $\gtrsim 110$ km s$^{-1}$. 
We measure proper motions of the large jet features identified by \citet{smi04} as a whole (shown as the larger boxes in Figure~\ref{fig:hh666_pm} and listed in Table~3 under the names given by \citealt{smi04}, e.g. HH~666~A) and the smaller knots that comprise each of those larger blobs separately (using the boxes shown as the inset images of each feature in Figure~\ref{fig:hh666_pm} and listed as, e.g. HH~666~A1, in Table~2). 
Both HH~666~M and O in the inner jet have high velocities. 
We expect that jet features near the driving source will have undergone the least amount of shock processing and interaction with the environment, and thus their velocities will be closest to the initial jet velocity. 
Velocities throughout HH~666~M remain remarkably consistent (see Figure~\ref{fig:all_pos_vel}a) despite its clumpy morphology. 
In contrast, knot velocities in HH~666~O appear to decrease as they move away from the driving source. 
Large proper motions and Doppler velocities from inner jet features indicate that the tilt of the jet away from the plane of the sky is $\alpha \approx 60^{\circ}$ (although the median inclination angle derived for all features in HH~666 is closer to 30$^{\circ}$, suggesting that the inclination for most of the jet may be somewhat smaller or may have changed over time). 
Combining the radial and transverse velocities yields total space velocities of $\sim 119-263$ km s$^{-1}$. 
Surprisingly, HH~666~E has the highest proper motions in the jet with a 3D space velocity of $263$ km s$^{-1}$, exceeding speeds in the inner jet and bow shocks by almost 100 km s$^{-1}$. 
\begin{table*}
\begin{minipage}{126mm}
\caption{Proper motions of small knots in HH~666}
\centering
\begin{tabular}{lrrrrrrr}
\hline\hline
Object & $\delta$x & $\delta$y & v$_T$$^{\mathrm{a}}$ & v$_R$$^{\mathrm{b}}$ & 
velocity$^{\mathrm{c}}$ & $\alpha$ & age$^{\mathrm{d}}$ \\
 & mas & mas & [km s$^{-1}$] & [km s$^{-1}$] & [km s$^{-1}$] & 
[degrees] & yr \\ 
\hline\hline
HH~666~A  &  $\overline{-49}(14)$  &  $\overline{-25}(33)$  &  $\overline{160}(39)$  &  -37(4)  &  $\overline{164}(38)$  &  $\overline{14}(4)$  &  $\overline{12869}(3553)$  \\
\hline
HH~666~A1  &  -55(25)  &  -34(25)  &  163(63)  &  ...  &  167(63)  &  13(5)  &  12114(4664)  \\
HH~666~A2  &  -64(25)  &  -54(25)  &  211(63)  &  ...  &  214(63)  &  10(3)  &  9183(2744)  \\
HH~666~A4  &  -55(25)  &  -39(50)  &  169(89)  &  ...  &  173(89)  &  12(6)  &  11651(6123)  \\
HH~666~A5  &  -51(12)  &  10(25)  &  132(33)  &  ...  &  137(33)  &  16(4)  &  14664(3697)  \\
HH~666~A6  &  -20(12)  &  -82(25)  &  212(61)  &  ...  &  215(61)  &  10(3)  &  9057(2627)  \\
HH~666~A7  &  -63(12)  &  12(37)  &  162(36)  &  ...  &  166(36)  &  13(3)  &  11853(2630)  \\
HH~666~A8  &  -38(12)  &  3(25)  &  97(32)  &  ...  &  104(32)  &  21(7)  &  20058(6579)  \\
HH~666~A9  &  -49(12)  &  -21(12)  &  134(31)  &  ...  &  139(32)  &  15(4)  &  14377(3382)  \\
\hline
HH~666~E  &  $\overline{-84}(11)$  &  $\overline{-30}(13)$  &  $\overline{227}(33)$  &  -130(4)  &  $\overline{262}(29)$  &  $\overline{30}(4)$  &  $\overline{6917(1004)}$  \\
\hline
HH~666~E1  &  -83(12)  &  -35(12)  &  226(32)  &  ...  &  261(32)  &  30(4)  &  6923(982)  \\
HH~666~E2  &  -84(25)  &  -39(12)  &  234(58)  &  ...  &  268(59)  &  29(6)  &  6656(1670)  \\
HH~666~E3  &  -98(25)  &  -35(25)  &  264(63)  &  ...  &  294(63)  &  26(5)  &  5837(1398)  \\
HH~666~E4  &  -72(25)  &  -10(25)  &  183(63)  &  ...  &  225(63)  &  35(9)  &  8253(2834)  \\
\hline
HH~666~M  &  $\overline{-66}(7)$  &  $\overline{8}(7)$  &  $\overline{169}(18)$  &  -190(4)  &  $\overline{255}(12)$  &  $\overline{48}(3)$  &  $\overline{1216}(410)$  \\
\hline
HH~666~M1  &  -72(25)  &  15(25)  &  186(63)  &  ...  &  266(63)  &  46(10)  &  1895(642)  \\
HH~666~M2  &  -69(12)  &  10(12)  &  176(32)  &  ...  &  259(32)  &  47(5)  &  1351(244)  \\
HH~666~M3  &  -62(12)  &  12(12)  &  159(32)  &  ...  &  248(32)  &  50(6)  &  1331(265)  \\
HH~666~M4  &  -66(25)  &  -6(25)  &  166(63)  &  ...  &  252(63)  &  49(11)  &  1102(417)  \\
HH~666~M5a  &  -75(25)  &  5(12)  &  189(63)  &  ...  &  268(63)  &  45(10)  &  773(257)  \\
HH~666~M7  &  -55(12)  &  11(25)  &  140(33)  &  ...  &  236(33)  &  54(6)  &  844(200)  \\
\hline
HH~666~O  &  $\overline{45}(10)$  &  $\overline{44}(16)$  &  $\overline{160}(44)$  &  210(4)  &  $\overline{266}(27)$  &  $\overline{53}(8)$  &  $\overline{2279}(1053)$  \\
\hline
HH~666~O1  &  57(25)  &  46(12)  &  184(53)  &  ...  &  279(53)  &  49(8)  &  1132(325)  \\
HH~666~O2  &  56(25)  &  66(25)  &  218(63)  &  ...  &  303(63)  &  44(8)  &  1373(395)  \\
HH~666~O3  &  42(25)  &  50(12)  &  165(47)  &  ...  &  267(47)  &  52(8)  &  2288(654)  \\
HH~666~O4  &  36(25)  &  23(12)  &  108(55)  &  ...  &  236(56)  &  63(12)  &  3628(1868)  \\
HH~666~O5  &  36(12)  &  35(25)  &  127(49)  &  ...  &  245(49)  &  59(10)  &  2973(1152)  \\
\hline
HH~666~N  &  $\overline{28}(28)$  &  $\overline{33}(9)$  &  $\overline{119}(46)$  &  93(4)  &  $\overline{153}(34)$  &  $\overline{40}(13)$  &  $\overline{6251}(2857)$  \\
\hline
HH~666~N1  &  -4(25)  &  26(50)  &  66(124)  &  ...  &  114(124)  &  55(51)  &  9529(17974)  \\
HH~666~N2  &  43(37)  &  43(17)  &  153(73)  &  ...  &  179(74)  &  31(12)  &  4293(2062)  \\
HH~666~N3  &  46(25)  &  29(37)  &  137(73)  &  ...  &  166(73)  &  34(14)  &  4929(2632)  \\
\hline
HH~666~I  &  $\overline{45}(14)$  &  $\overline{27}(11)$  &  $\overline{136}(26)$  &  67(4)  &  $\overline{152}(24)$  &  $\overline{27}(4)$  &  $\overline{9320}(1926)$  \\
\hline
HH~666~I1  &  37(25)  &  34(25)  &  128(63)  &  ...  &  144(63)  &  28(12)  &  8195(4026)  \\
HH~666~I2  &  20(25)  &  36(25)  &  104(63)  &  ...  &  123(63)  &  33(16)  &  10052(6070)  \\
HH~666~I3  &  40(25)  &  43(25)  &  147(63)  &  ...  &  162(63)  &  25(9)  &  7384(3152)  \\
HH~666~I4  &  73(25)  &  25(25)  &  195(63)  &  ...  &  206(63)  &  19(6)  &  5966(1924)  \\
HH~666~I5  &  48(25)  &  27(12)  &  140(57)  &  ...  &  155(57)  &  26(9)  &  8487(3457)  \\
HH~666~I6  &  50(25)  &  10(12)  &  128(62)  &  ...  &  144(62)  &  28(11)  &  10322(4998)  \\
HH~666~I7  &  58(25)  &  14(25)  &  150(63)  &  ...  &  164(63)  &  24(9)  &  9068(3800)  \\
HH~666~I8  &  42(25)  &  17(37)  &  116(68)  &  ...  &  134(68)  &  30(15)  &  11463(6733)  \\
HH~666~I9  &  38(12)  &  21(12)  &  110(31)  &  ...  &  129(32)  &  31(7)  &  12487(3564)  \\
HH~666~I10  &  40(25)  &  40(25)  &  144(63)  &  ...  &  159(63)  &  25(10)  &  9778(4265)  \\
\hline
HH~666~C  &  $\overline{70}(18)$  &  $\overline{20}(12)$  &  $\overline{185}(45)$  &  67(4)  &  $\overline{197}(43)$  &  $\overline{21}(4)$  &  $\overline{13679}(2714)$  \\
\hline
HH~666~C1  &  63(12)  &  11(37)  &  161(35)  &  ...  &  175(35)  &  23(5)  &  14477(3141)  \\
HH~666~C2  &  57(12)  &  5(37)  &  143(32)  &  ...  &  158(33)  &  25(5)  &  16356(3703)  \\
HH~666~C3  &  87(37)  &  29(25)  &  230(91)  &  ...  &  240(91)  &  16(6)  &  10149(4031)  \\
HH~666~C4  &  56(25)  &  16(25)  &  148(63)  &  ...  &  163(63)  &  24(9)  &  16087(6820)  \\
HH~666~C5  &  62(37)  &  12(25)  &  161(93)  &  ...  &  174(93)  &  23(12)  &  15093(8731)  \\
HH~666~C7  &  70(25)  &  40(25)  &  204(63)  &  ...  &  215(63)  &  18(5)  &  12371(3810)  \\
HH~666~C8  &  105(12)  &  14(12)  &  269(32)  &  ...  &  277(32)  &  14(2)  &  9388(1132)  \\
HH~666~C9  &  56(12)  &  32(12)  &  163(32)  &  ...  &  176(32)  &  22(4)  &  15512(3014)  \\
\hline
\end{tabular}
\begin{tabular}{l}
Proper motions measured for small knots within the larger features in HH~666 (small boxes on inset \\ 
images in Figure~\ref{fig:hh666_pm}). We also report the mean and standard deviation of the measurements within each \\ large feature. 
Radial velocities for each feature are those measured for the larger jet feature (see Table~3 \\ and \citealt{smi04}). 
$^a$ The transverse velocity, assuming a distance of 2.3 kpc. 
$^b$ The radial velocity \\ measured from spectra. 
$^c$ The total space velocity, assuming the average inclination measured for other \\ knots in the jet 
when a radial velocity is not measured directly. 
$^d$ Time for the object to reach its current \\ position at the measured velocity, assuming ballistic motion. \\
\end{tabular}
\end{minipage}
\label{t:hh666_pm}
\end{table*}

\begin{figure*}
\centering
$\begin{array}{ll}
\includegraphics[angle=0,scale=0.55]{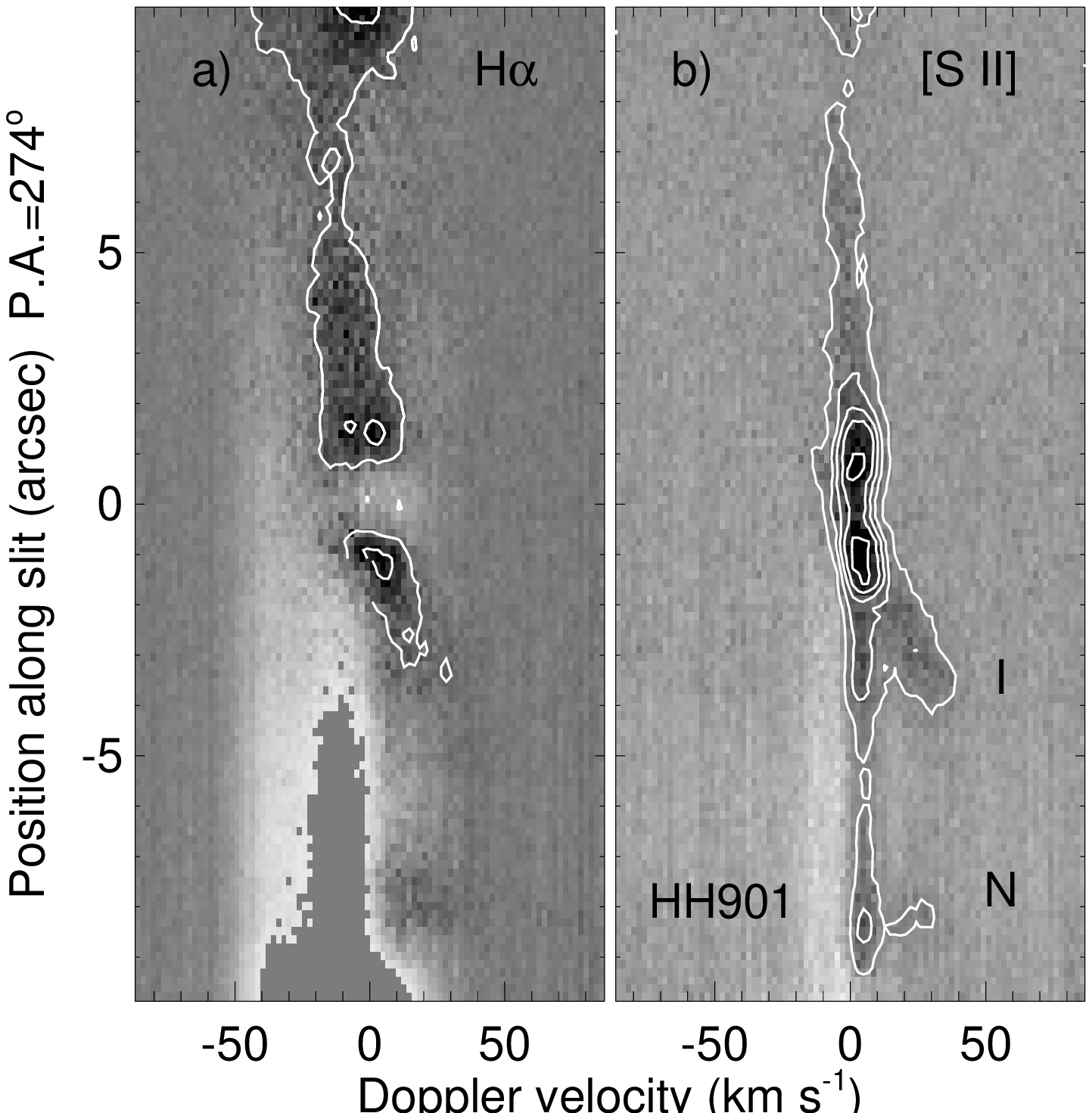} & 
\includegraphics[trim=0mm -22mm 0mm 30mm,angle=0,scale=0.385]{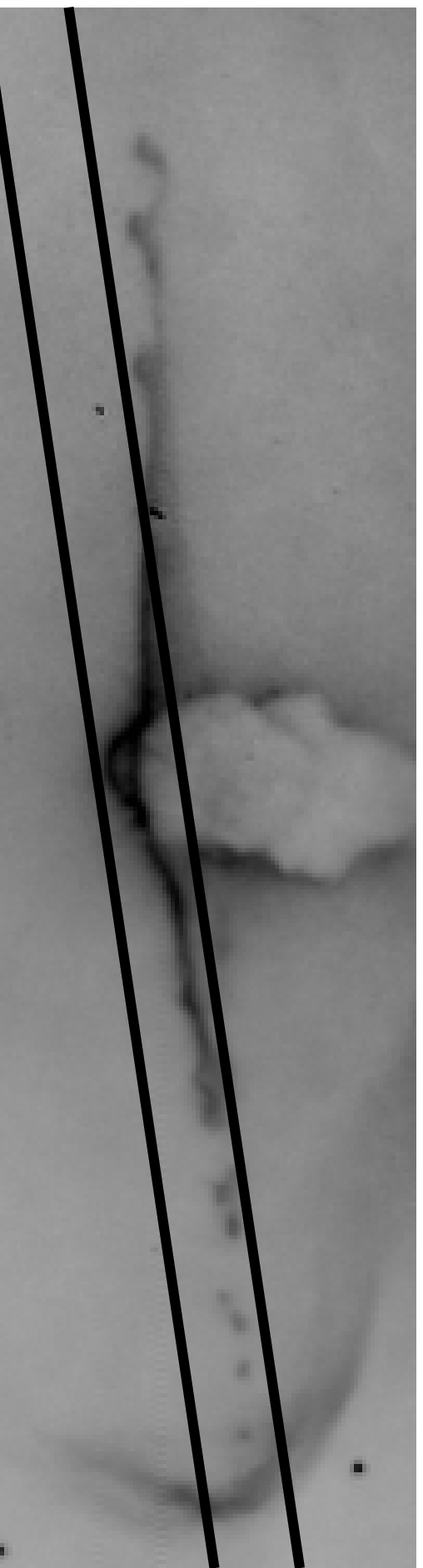} \\
\end{array}$
\caption{Position-velocity diagrams of HH~901 in (a) H$\alpha$, (b) [S~II] (the $\lambda6717$ and $\lambda6731$ lines combined), with an image of the slit position with approximately the same spatial scale shown to the right. Emission from the background has been suppressed but not completely extracted. Contours are not flux calibrated. }\label{fig:hh901_spec} 
\end{figure*}

\textit{HH~901:} 
Blobs in HH~901 move along the jet axis, culminating in the eastern and western bow shocks (see Figure~\ref{fig:hh901_pm}). 
Measuring the proper motions of the apex and wing of the eastern bow shock separately confirms that the bow tip moves faster than the skirt as is commonly seen in HH jets \citep{rei01}. 
Optical spectra of the western limb of HH~901 show some evidence of a Hubble flow from the inner jet (particularly in [S~II]) and high velocity emission from the terminal bow shock. 
Radial velocities from the western portion of the jet (the misaligned eastern limb falls outside of the slit) are $\approx 20$ km s$^{-1}$. 
Since the transverse velocities are $\sim 3\times$ the Doppler velocities, this implies that the tilt angle of the relative to the plane of the sky is $\alpha \sim 10-20^{\circ}$ (see Figure~\ref{fig:hh901_spec}). 


\begin{figure*}
\centering
$\begin{array}{ll}
\includegraphics[angle=0,scale=0.55]{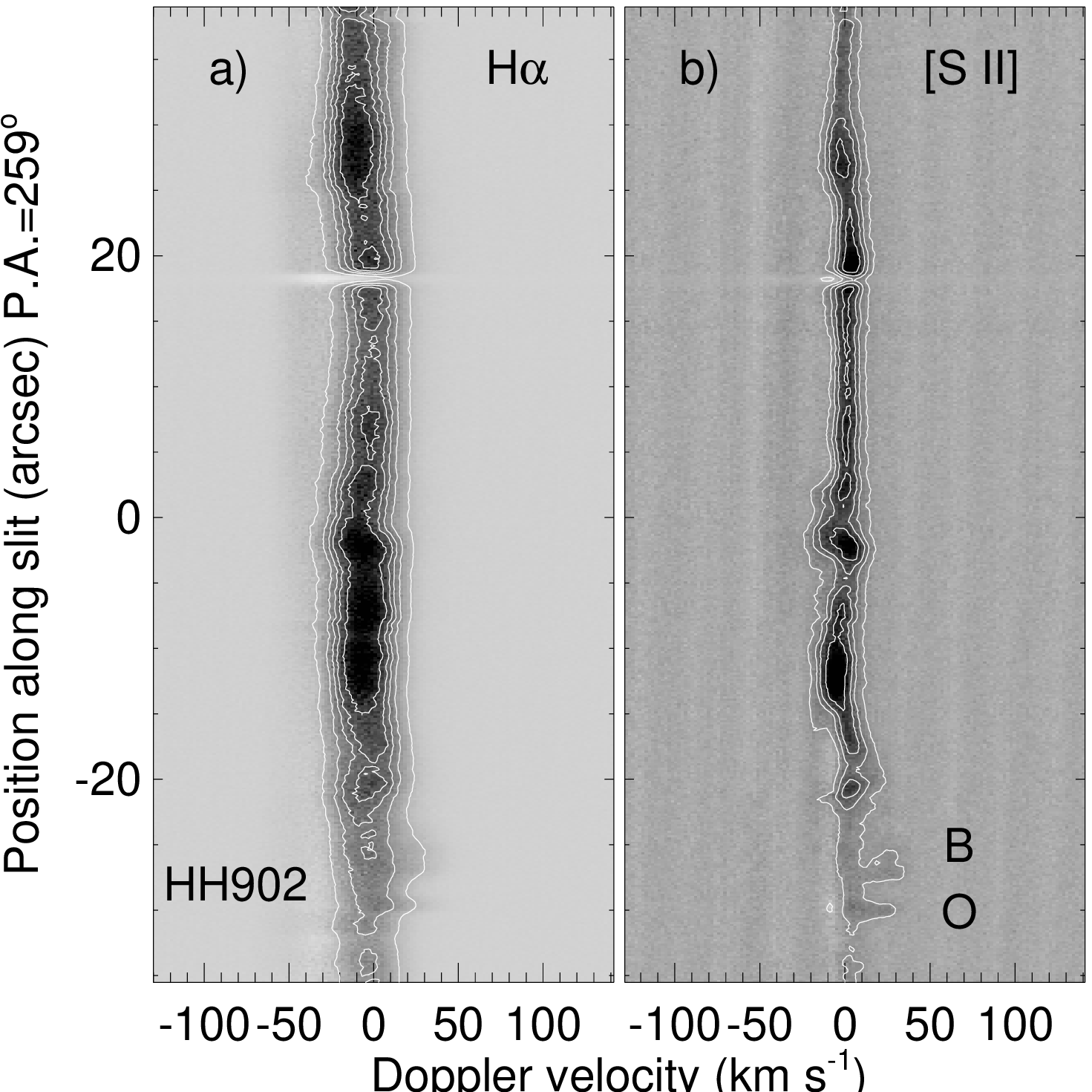} &
\includegraphics[trim=0mm -33.5mm 0mm 50mm,angle=0,scale=0.255]{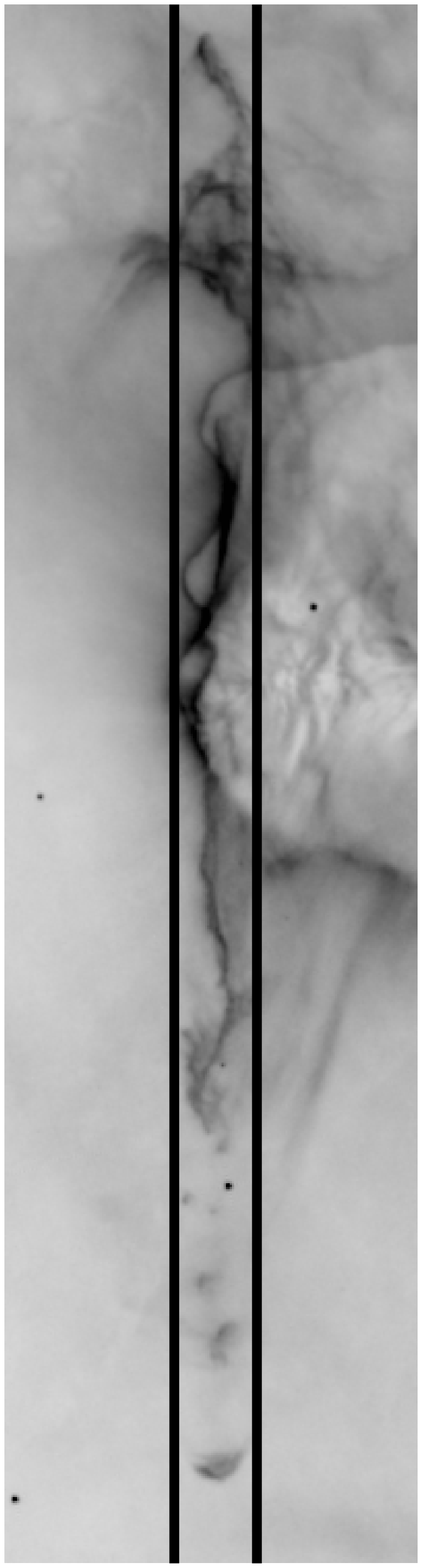} \\
\end{array}$
\caption{Long-slit spectra along the HH~902 jet axis. As in Figure~\ref{fig:hh901_spec}, panel (a) shows H$\alpha$, (b) a combination of the $\lambda6717$ and $\lambda6731$ [S~II] lines, and an image of the jet with approximately the same spatial resolution showing the slit position is plotted to the right. Contours are not flux calibrated. The horizontal white stripe in both spectra is a gap in the CCD detector.}\label{fig:hh902_spec} 
\end{figure*}

\textit{HH~902:} 
The brightest knots in HH~902 lie along a single jet axis, with more diffuse features deviating from this axis as though they have been blown back by feedback from Trumpler 14 \citep[see Figure~\ref{fig:hh902_pm} and][]{smi10}. 
Proper motions of the H$\alpha$ blobs in HH~902 are directed along this axis with no evidence for significant motions perpendicular to it (our precision indicates that perpendicular velocities are $\lesssim 50$ km s$^{-1}$). 
[Fe~{\sc ii}] emission identified by \citet{rei13} lies along the jet axis defined by the proper motions, demonstrating that [Fe~{\sc ii}] does indeed trace the inner jet and not a photodissociation region (PDR) on the pillar surface. 

\citet{ohl12} proposed that an IR source (J104401.8–593030) near the tip of the HH~902 pillar drives the jet. 
This same source can be seen in our H$\alpha$ images from \textit{HST} and lies more than 2\arcsec\ away from the jet axis. 
Proper motions demonstrate that it cannot be the protostar driving the HH~902 outflow. 
This suggests caution when inferring position-based associations of jets and driving sources using low-resolution mid/far-IR images alone. 

Like HH~901, the HH~902 flow axis lies close to the plane of the sky. 
The highest radial velocity in the optical spectra, from the redshifted western limb of the jet, is $\approx 25$ km s$^{-1}$, although emission from the rest of the jet is confined to $|v_R| \la 15$ km s$^{-1}$. 
Small radial velocities in the optical spectra combined with higher transverse speeds from the proper motions constrain the tilt of the jet axis from the plane of the sky to be $\alpha \sim 10-25^{\circ}$ (see Figure~\ref{fig:hh902_spec}).

\textit{HH~1066:} 
Unlike the other HH jets studied here, HH~1066 does not have a clear bipolar jet structure in H$\alpha$ images, leading \citet{smi10} to identify it as a candidate jet. 
However, near-IR [Fe~{\sc ii}] emission clearly traces a bright, collimated jet emerging from the apex of the globule, as well as a western bow shock \citep{rei13}. 
Faint arc-like H$\alpha$ emission that is confused with the pillar edge in single-epoch images is clearly seen to be the eastern bow shock in an H$\alpha$ difference image because of its high proper motion (see Figure~\ref{fig:hh1066}b).
Neither the eastern nor western bow shock lies along the same axis as bright [Fe~{\sc ii}] emission from the inner jet \citep{rei13}. 
The eastern bow shock is moving at $\sim 215$ km s$^{-1}$ parallel to, but offset $\sim 0\farcs 3$ southwest of this axis. 
In contrast, the western bow shock moves almost due west with a velocity of $\sim 140$ km s$^{-1}$, and is offset to the south by an angle $\sim 20^{\circ}$ relative to the [Fe~{\sc ii}] jet axis. 
Taken together, the offset of the bow shocks suggests that the jet axis is bent \textit{toward} Trumpler 14. 
Possible causes of this asymmetry are discussed in Section~\ref{ss:bending}. 

Two new jet blobs, similar to those discovered close to the driving sources of HH~668 in Orion \citep{smi05} and DG~Tau \citep{aa11}, are apparent in the H$\alpha$ difference image of HH~1066 where the knots can be seen moving along the flow axis amid the more diffuse emission of the inner jet. 
These two knots, are both $\sim 0.5$\arcsec\ away from the presumed position of the driving source (between the bright spots where the jet emerges from the globule, see Figure~\ref{fig:hh1066}c). 
While they lie too close to the bright edges of the globule for reliable estimates of their proper motions (see Figure~\ref{fig:hh1066}), an intensity tracing along the jet shows an offset of $\sim 0\farcs1$ between the two epochs for the western blob, indicating transverse velocities of at least 250 km s$^{-1}$ (see Figure~\ref{fig:hh1066}d). 
This suggests a dynamical age of $\sim 35$ years for the two newly revealed knots. 
Additional epochs will be required for more precise constraints on their proper motions.  

\begin{figure}
\centering
$\begin{array}{cc}
\includegraphics[angle=0,scale=0.315]{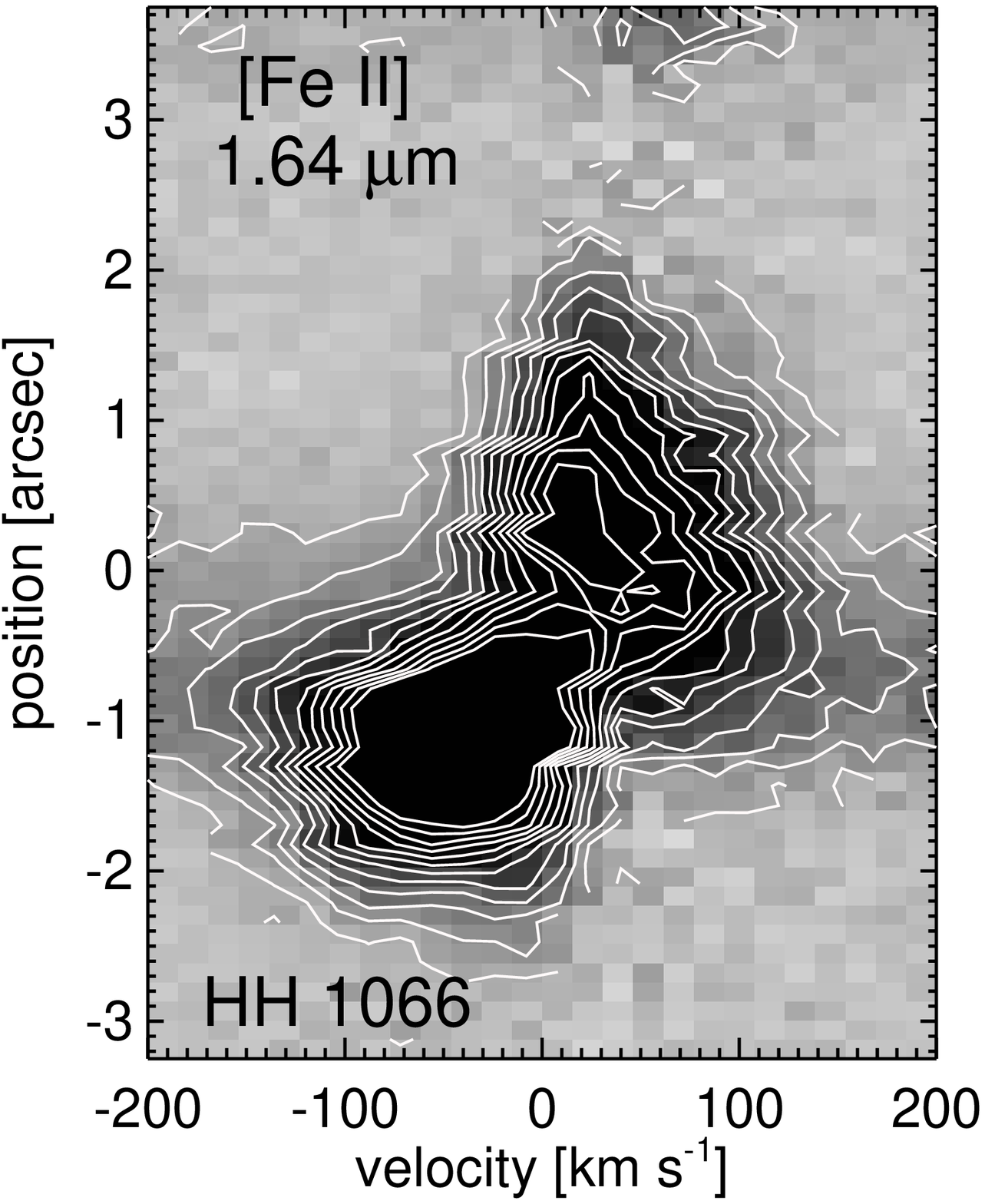} & 
\includegraphics[trim=0mm -31mm 0mm 0mm,angle=0,scale=0.3265]{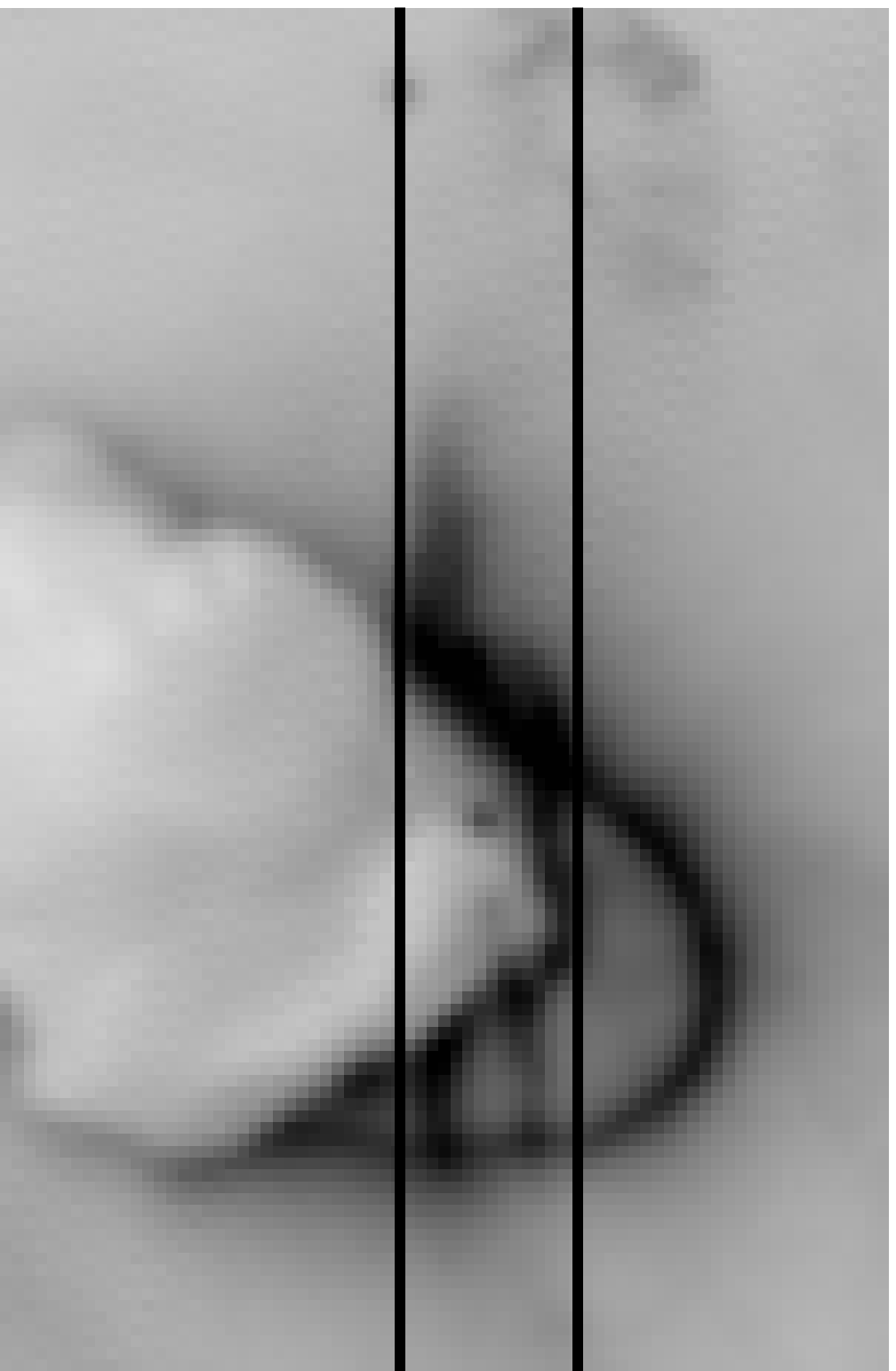} \\
\end{array}$
\caption{Position-velocity diagram of HH~1066 in [Fe~{\sc ii}] 1.644 \micron\ with contours overplotted to show the distribution of flux shown alongside an image with approximately the same spatial resolution with the slit position superposed. Continuum from the driving source separates the redshifted (western) and blueshifted (eastern) sides of the jet. 
}\label{fig:hh1066_spec}
\end{figure}

The likely driving source of HH~1066 \citep{ohl12,rei13} lies within $\sim 0\farcs25$ of the bright spots where the jet emerges from the globule (with the discrepancy attributable to \textit{Spitzer's} angular resolution). 
Near-IR spectra of the [Fe~{\sc ii}] 1.64 \micron\ line from HH~1066 trace Doppler velocities up to $\pm 100$ km s$^{-1}$ from the inner jet and western bow shock (see Figure~\ref{fig:hh1066_spec}). 
The range of radial velocities is especially large near the jet base, in both the redshifted and blueshifted sides of the jet. 
Unlike the narrow range of velocities traced by a collimated jet that is slowed in a shock, the broad range of velocities are more reminiscent of the conical outflow structures seen in, e.g. the Orion BN/KL outflow \citep{gre98}. 
Comparing the approximate velocity of the recently ejected knots with the maximum radial velocity in the [Fe~{\sc ii}] 1.64 \micron\ line constrains the tilt angle to be $\alpha \sim 20-30^{\circ}$. 

%


\section{Discussion}\label{s:discussion}
 
\subsection{Velocity structure}\label{ss:structure}
\begin{figure}
\centering
\includegraphics[angle=90,scale=0.35]{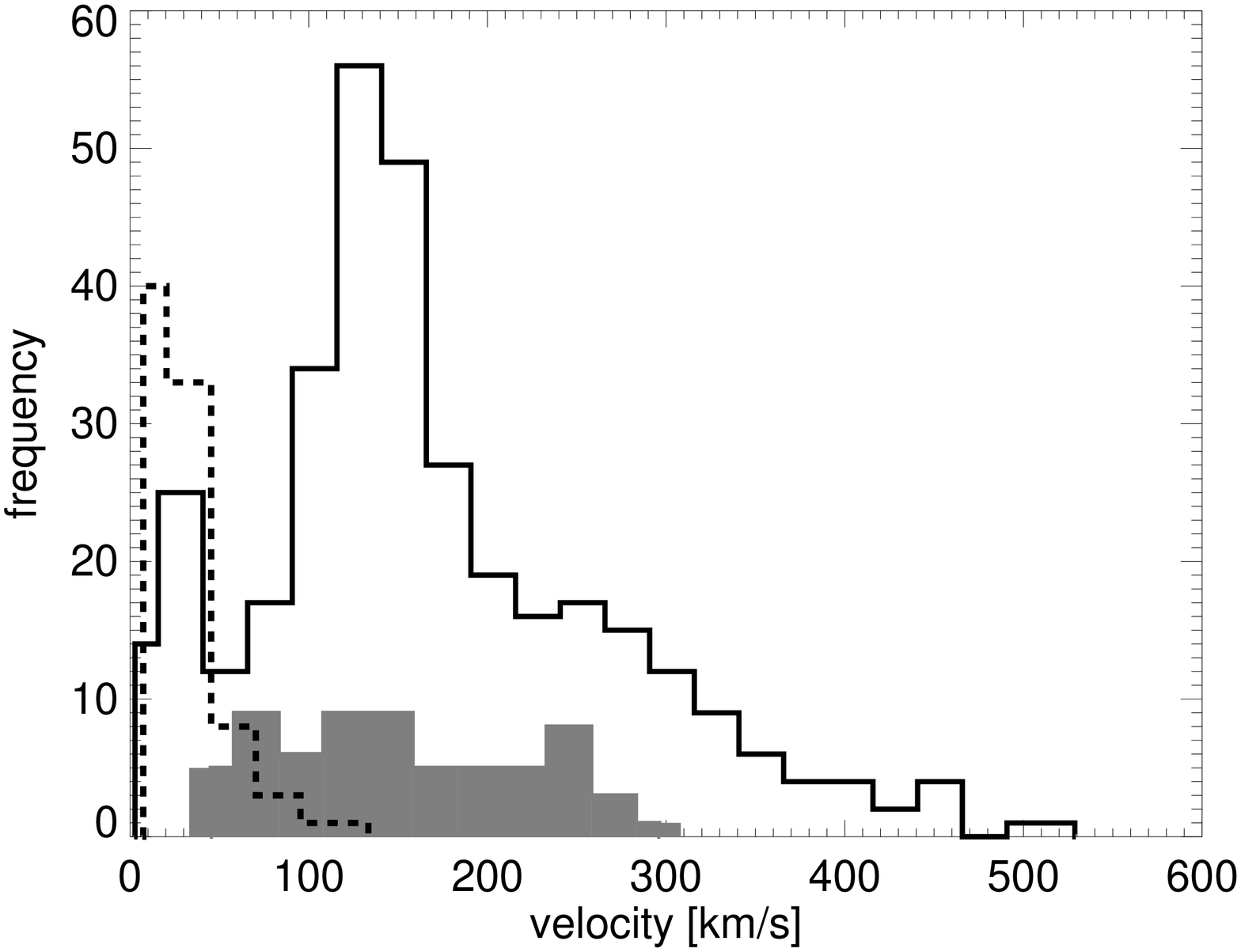}
\caption{Histogram of the total space velocities measured in other jets (mostly associated with low-mass protostars). The solid line shows velocities measured in other HH jets, the dotted line shows H$_2$ jets, and the shaded histogram shows the velocities measured for the various knots in HH~666, HH~901, HH~902, and HH~1066. 
The references for the HH jet velocities from the literature are as follows: 
\citet{bal02}, \citet{bal12}, \citet{dev97}, \citet{dev09}, \citet{har01}, 
\citet{har05}, \citet{har07}, \citet{kaj12}, \citet{mcg07}, \citet{nor01}, 
\citet{rei02}, \citet{smi05}, and \citet{yus05}. 
H$_2$ jet velocities are from \citet{zha13}. 
}\label{fig:vhist} 
\end{figure}

The proper motion measurements of HH~666, HH~901, HH~902, and HH~1066 combined with Doppler shifts provide the full three-dimensional space motions between the first and second epochs for these four HH jets in the Carina nebula. 
Measured velocities from the Carina jets are similar to speeds measured in the outflows driven by low-luminosity protostars \citep[$\sim 100-200$ km s$^{-1}$, see Figure~\ref{fig:vhist} and][]{rei01}, somewhat slower than the high velocity outflows inferred from blueshifted forbidden line emission in spectra of some Herbig Ae/Be stars \citep{cor97}. 
However, we measure velocities in these spatially resolved outflows far from the driving source, where interaction with the environment may have altered the measured outflow velocity from the launch speed (more significant for jets from embedded protostars).
All four jets show velocities $\ga 100$ km s$^{-1}$. 
Many knots are slower than the high velocities found throughout HH~399 \citep{yus05}, an irradiated outflow in the Trifid with a high measured mass-loss rate similar to the inferred mass-loss rate for these four jets in Carina \citep[$\sim 10^{-6}$ M$_{\odot}$ yr$^{-1}$, see][]{rei13}. 

Combining total space velocities with the high densities required for the survival of Fe$^{++}$ \citep[$n \ga 3.4 \times 10^4$ cm$^{-3}$, see][]{rei13}, we find mass-loss rates of a few $\times 10^{-6}$ M$_{\odot}$ yr$^{-1}$ for all four jets. 
These high mass-loss rates exceed the \textit{accretion} rate of the two HH jets driven by intermediate-mass protostars studied by \citet{ell13} and are more than two orders of magnitude larger than the average mass-loss rates measured from T Tauri stars \citep{har95}. 
Interestingly, this is comparable to the observed mass-loss rates in FU Orionis outbursts \citep{hc95}. 
Better constraints on the density of the HH jets in Carina from spectroscopy will further improve the mass-loss rates (Reiter et al., in preparation).

\begin{figure*}
\centering
$\begin{array}{cccc}
\includegraphics[trim=10mm 0mm 10mm 0mm,angle=0,scale=0.235]{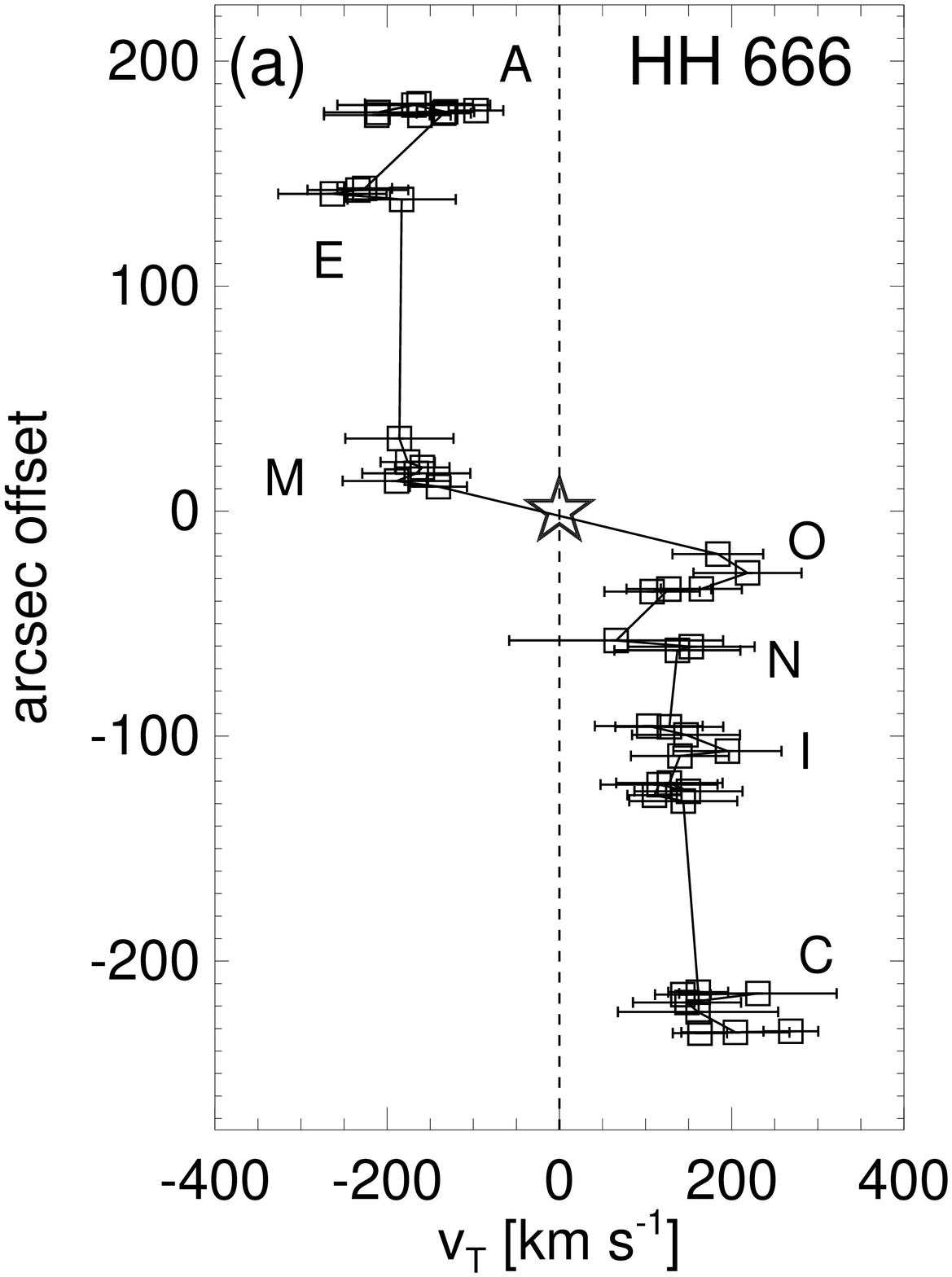} &
\includegraphics[trim=10mm 0mm 10mm 0mm,angle=0,scale=0.235]{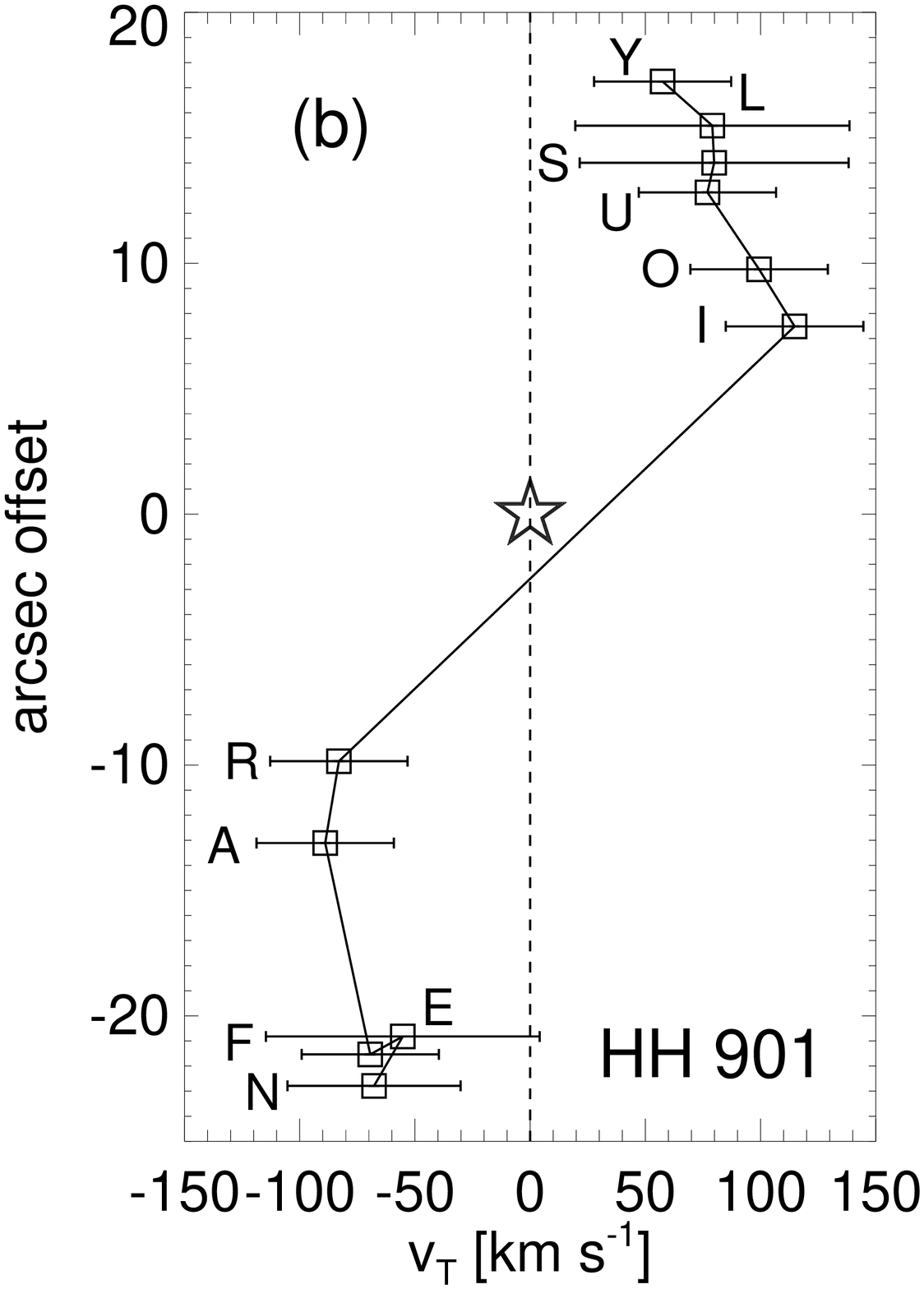} & 
\includegraphics[trim=10mm 0mm 10mm 0mm,angle=0,scale=0.235]{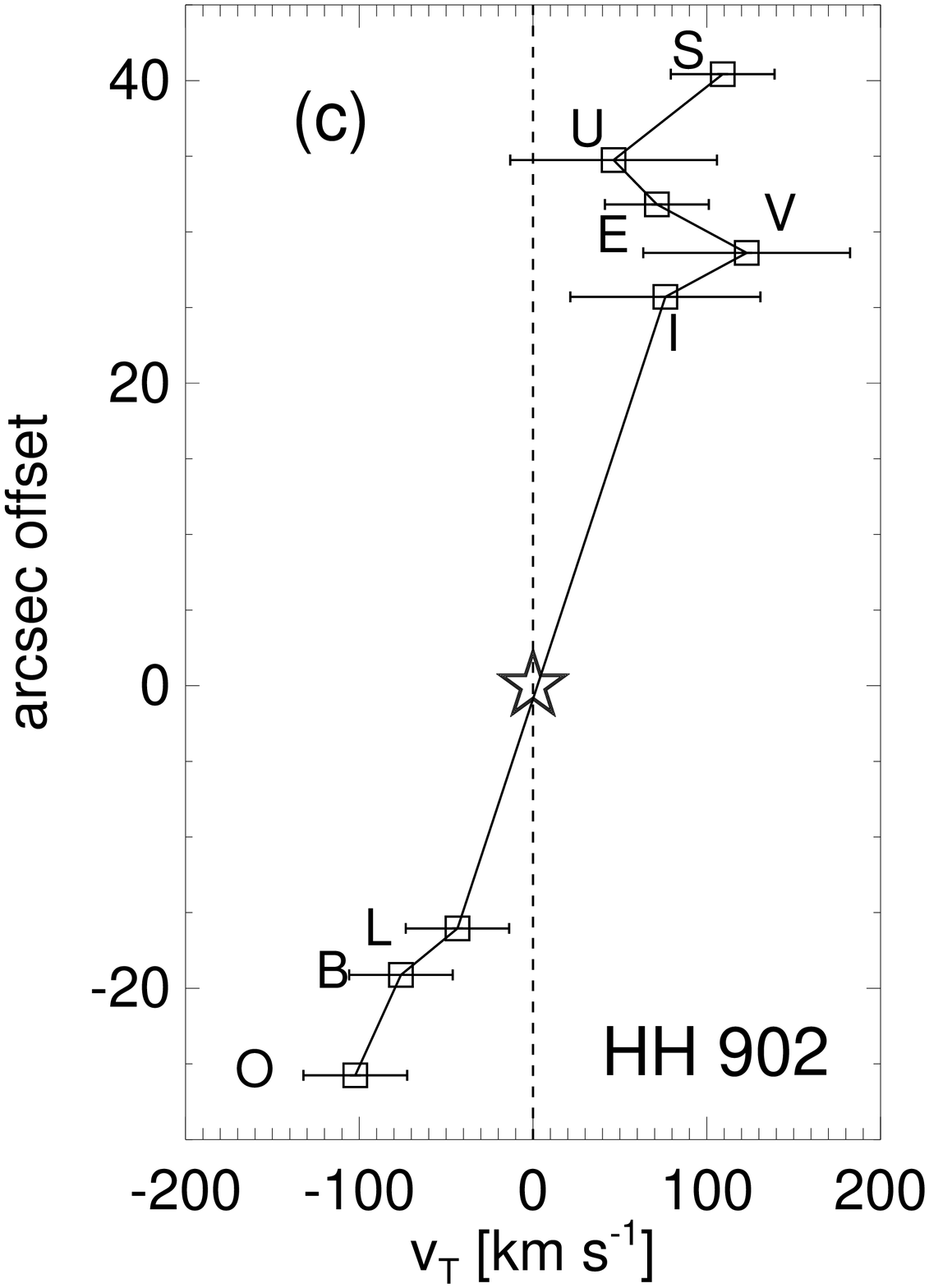} &
\includegraphics[trim=10mm 0mm 10mm 0mm,angle=0,scale=0.235]{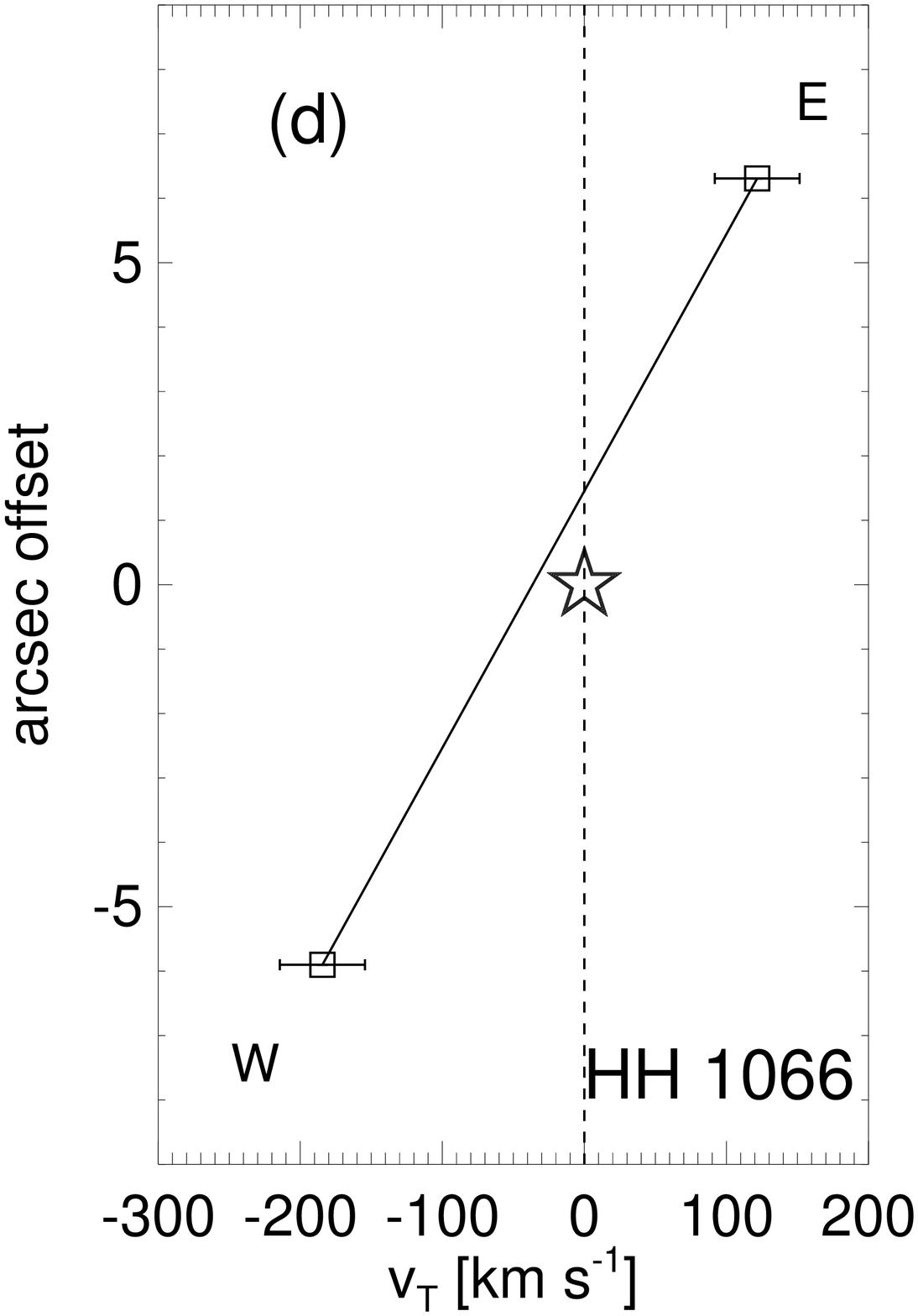} \\
\end{array}$
\caption{Position-velocity plots showing the transverse velocities measured for (a) HH~666, (b) HH~901, (c) HH~902, and (d) HH~1066 (proper motions of the driving sources, if any, are below our detection limit and assumed to be zero). The highest velocities are usually found in the bow shocks and/or the innermost knots. 
}\label{fig:all_pos_vel} 
\end{figure*}

None of the jets show evidence of a velocity decrease $\gtrsim 100$ km s$^{-1}$ with increasing distance from the driving source (see Figure~\ref{fig:all_pos_vel}) as \citet{dev97} observe in HH~34. 
\citet{dev97} suggest that the monotonic decrease in velocity observed in HH~34 may result from a young outflow interacting with a surrounding medium that has not been accelerated by momentum transfer from previous ejecta. 
In an H~{\sc ii} region with a temperature of 10,000 K and a density of 100 cm$^{-3}$ (the approximate H~{\sc ii} region density near HH~666 estimated from the Str\"{o}mgren condition; the density near HH~901, HH~902, and HH~1066 is expected to be nearly an order of magnitude higher, see Section~\ref{ss:bending}), the ambient pressure is $\sim 1.5 \times 10^{-10}$ dynes cm$^{-2}$. 
In contrast, the ram pressure of an individual jet knot with the high densities we find in Carina is orders of magnitude higher; for a density $n_e \gtrsim 10^4$ cm$^{-3}$ \citep{rei13}, and velocity of 100 km s$^{-1}$, the ram pressure is $\sim 1.5 \times 10^{-8}$ dynes cm$^{-2}$. 
Outside the parent pillar, few forces will act to decelerate the jet, with the possible exception of the jet colliding with a nearby high-density pillar (as may be the case for HH~1066, see Sections~\ref{s:results} and \ref{ss:bending}).

\begin{table*}
\begin{minipage}{126mm}
\caption{Proper motions measured for various jet features}
\centering
\begin{tabular}{lrrrrrrr}
\hline\hline
Object & $\delta$x & $\delta$y & v$_T$$^{\mathrm{a}}$ & v$_R$$^{\mathrm{b}}$ & 
velocity$^{\mathrm{c}}$ & $\alpha$ & age$^{\mathrm{d}}$ \\
 & mas & mas & [km s$^{-1}$] & [km s$^{-1}$] & [km s$^{-1}$] & 
[degrees] & yr \\ 
\hline\hline
\multicolumn{8}{c}{\textbf{HH~666$^{\star}$} ($\Delta t = 4.32$ yr)} \\ 
\hline
HH~666~D  &  ...  &  ...  &  ...  &  -11  &  22  &  30$^{\dagger}$  &  45110  \\
\hline
HH~666~A  &  -50(25)  &  -13(25)  &  130(63)  &  -37(4)  &  135(63)  &  16(7)  &  14957(7210)  \\
HH~666~E  &  -84(12)  &  -33(12)  &  229(32)  &  -130(4)  &  263(32)  &  30(3)  &  6735(945)  \\
HH~666~M  &  -54(12)  &  7(25)  &  136(32)  &  -190(4)  &  234(33)  &  54(6)  &  1364(324)  \\
HH~666~O  &  41(12)  &  27(12)  &  124(31)  &  210(4)  &  244(32)  &  60(6)  &  3045(776)  \\
HH~666~N  &  9(50)  &  28(37)  &  75(97)  &  93(4)  &  119(97)  &  51(36)  &  8719(11340)  \\
HH~666~Ib  &  32(25)  &  30(25)  &  110(63)  &  67(4)  &  129(63)  &  31(15)  &  11358(6458)  \\
HH~666~Ia  &  38(25)  &  21(25)  &  109(63)  &  67(4)  &  128(63)  &  32(15)  &  12488(7179)  \\
HH~666~C  &  63(37)  &  6(25)  &  159(94)  &  67(4)  &  172(94)  &  23(12)  &  15304(9058)  \\
\hline
\multicolumn{8}{c}{\textbf{HH~901} ($\Delta t = 4.55$ yr)} \\ 
\hline
HH~901~Y  &  25(12)  &  -13(25)  &  68(38)  &  ...  &  70(38)  &  ...  &  3670(2037)  \\
HH~901~L  &  23(25)  &  -2(12)  &  55(59)  &  ...  &  57(59)  &  ...  &  4111(4420)  \\
HH~901~S  &  29(12)  &  3(12)  &  69(30)  &  ...  &  72(30)  &  ...  &  3387(1456)  \\
HH~901~U  &  37(12)  &  1(12)  &  89(30)  &  ...  &  92(30)  &  ...  &  1608(540)  \\
HH~901~O  &  34(12)  &  -4(12)  &  83(30)  &  ...  &  86(30)  &  ...  &  1295(466)  \\
HH~901~I  &  -47(12)  &  -10(12)  &  115(30)  &  20(4)  &  119(30)  &  10(3)  &  712(186)  \\
HH~901~R  &  -40(12)  &  -10(12)  &  99(30)  &  ...  &  103(30)  &  ...  &  1072(323)  \\
HH~901~A  &  -29(12)  &  -13(12)  &  77(30)  &  ...  &  80(30)  &  ...  &  1818(705)  \\
HH~901~F  &  -32(25)  &  -7(12)  &  80(58)  &  ...  &  83(58)  &  ...  &  1914(1399)  \\
HH~901~E  &  -33(25)  &  3(25)  &  79(60)  &  ...  &  82(60)  &  ...  &  2137(1609)  \\
HH~901~N  &  -21(12)  &  -11(12)  &  57(30)  &  23(4)  &  60(30)  &  22(11)  &  3273(1696)  \\
\hline
\multicolumn{8}{c}{\textbf{HH~902} ($\Delta t = 4.55$ yr)} \\ 
\hline
HH~902~S  &  43(12)  &  1(12)  &  102(30)  &  ...  &  107(30)  &  ...  &  2747(803)  \\
HH~902~U  &  31(12)  &  4(12)  &  76(30)  &  ...  &  80(30)  &  ...  &  2744(1077)  \\
HH~902~E  &  18(12)  &  -4(12)  &  43(30)  &  ...  &  46(30)  &  ...  &  4027(2759)  \\
HH~902~V  &  -28(25)  &  14(12)  &  76(55)  &  ...  &  80(55)  &  ...  &  3681(2644)  \\
HH~902~I  &  -44(25)  &  27(25)  &  123(60)  &  ...  &  129(60)  &  ...  &  2539(1230)  \\
HH~902~L  &  -30(12)  &  1(25)  &  71(30)  &  ...  &  75(30)  &  ...  &  4868(2042)  \\
HH~902~B  &  -19(25)  &  5(25)  &  46(59)  &  23(4)  &  49(59)  &  26(27)  &  8177(10495)  \\
HH~902~O  &  -44(12)  &  11(12)  &  109(30)  &  18(4)  &  115(30)  &  9(4)  &  4040(1107)  \\
 \hline
\multicolumn{8}{c}{\textbf{HH~1066} ($\Delta t = 4.55$ yr)} \\ 
\hline
HH~1066~W  &  -46(12)  &  -20(12)  &  122(30)  &  74(15)  &  142(30)  &  31(8)  &  565(139)  \\
HH~1066~E  &  55(12)  &  53(12)  &  184(30)  &  ...  &  216(30)  &  ...  &  349(57)  \\
\hline
\end{tabular}
\medskip 
\footnotesize 
\begin{tabular}{l}
Uncertainties for each quantity are listed in parenthesis. \\
$^a$ The transverse velocity, assuming a distance of 2.3 kpc. \\
$^b$ The radial velocity. \\
$^c$ The total space velocity, assuming the average inclination measured for other knots in the jet \\ 
when a radial velocity is not measured directly. \\
$^d$ Time for the object to reach its current position at the measured velocity, assuming ballistic motion. \\
$^{\star}$ Proper motions for HH~666 are measured for the large blobs as a whole (e.g. HH~666~A, large boxes \\ 
in Figure~\ref{fig:hh666_pm}). 
The tilt angle $\alpha$ is calculated using the radial velocities reported by \citet{smi04} \\ 
$^{\dagger}$ except for HH~666~D, where we assume $\alpha=30^{\circ}$. \\
\end{tabular}
\end{minipage}
\label{t:pm}
\end{table*}

\subsection{Jet bending}\label{ss:bending}
\citet{rei13} considered the possibility that HH~901 may be bent away from Trumpler 14 by the rocket effect due to the combined UV radiation of many nearby O-type stars. 
Photoionizing only the side of the jet facing the O-stars creates a thrust that pushes the jet in the opposite direction of the ionizing source. 
With the discovery of the eastern bow shock of HH~1066, it appears that unlike HH~901, HH~1066 bends \textit{toward} Trumpler 14. 
Even though both jets reside within the same cloud complex, they are affected by different amounts of massive star feedback and propagate into somewhat different environments. 
HH~901 emerges from the apex of the cloud complex, whereas HH~1066 emerges from a small globule that appears to lie just in front of the heavily irradiated walls of the larger cloud. 
Local feedback on HH~1066 may be dominated by pressure from the photoevaporative flow coming off the face of the molecular cloud, deflecting the jet \textit{toward} Trumpler 14 \citep{bal01,smi10}. 
Uncertainties in the 3D geometry prevent detailed calculation of the two competing pressures, and thus it remains unclear whether the inward force of the plasma flowing off the irradiated molecular surfaces behind the jet overwhelms the outward force from the rocket effect.  

Alternate explanations for the C-shaped bend in HH~1066 include the outward motions of the HH~1066 driving source \citep[as][argued from the C-symmetry of the bent HH jets in NGC~1333]{bal01}. 
This requires that the driving source would have had to move $\sim 0.5$\arcsec\ within the short dynamical time ($\sim 450$ yrs) implied by the high velocity bow shocks. 
We note that the implied proper motion velocity, $\sim 12$ km s$^{-1}$, is suspiciously similar to the speed of the photoevaporative flow coming off the nearby molecular cloud that may also deflect the jet. 
An oblique collision of the outflow with the neighboring molecular cloud could also create the apparent offset from the jet axis \citep[e.g. HH~110/270,][]{rei96,rag02}. 
Radial velocities in the HH~1066 position-velocity diagram indicate that the western limb of the jet is redshifted, possibly propagating toward the pillar behind it. 
Finally, the complicated morphology of HH~1066 may reflect the combined effects of two separate outflows from the formation of a binary pair. 
A pair of orthogonal outflows viewed at the estimated tilt angle of $\sim 45^{\circ}$ away from the plane of the sky may be confused in the lower resolution infrared observations that trace the jet[s] inside the optically thick dust pillar \citep[see][]{rei13}. 
Different driving sources may also account for the discrepancy in the age and velocity of the two bow shocks as the two components of the binary will not necessarily have synchronised outflow behavior. 
For example, IRAS 16293-2422, shows a quadrupolar outflow structure on large scales, but only source A appears to be actively driving an outflow when observed at high resolution \citep{cha05,yeh08,kri13}. 

\subsection{Shocks}\label{ss:shocks}
Ejection velocities in protostellar jets may be time-variable (see Section~\ref{ss:variability}), producing internal shocks. 
These internal working surfaces will have velocities comparable to the amplitude of the velocity perturbation \citep[e.g.][]{rag92}. 
For jets that lie near the plane of the sky (as the jets in Carina do, see Section~\ref{s:results}), the difference between the inclination-corrected proper motions of adjacent knots can be used to estimate the shock velocity at which knots collide. 
We find a range of velocity jumps from $\sim 10-150$ km s$^{-1}$, although the uncertainty of the measured velocities leads to substantial uncertainty in each individual shock velocity (see Table~3). 
Fast, dissociative J-shocks ($v_{shock} > 30-40$ km s$^{-1}$) are expected to be bright in near-IR [Fe~{\sc ii}] lines \citep{nis02,pod06}. 
Indeed, near-IR imaging reveals bright [Fe~{\sc ii}] emission from clumps throughout HH~666 \citep[see][]{rei13} where velocity differences between jet features are also high, leading to higher velocity shocks.

\subsection{Ages}\label{ss:ages} 
Proper motions give a representative dynamical age of each feature. 
Shocks will decelerate the jet, leading to slower observed velocities, and ages that overestimate the true age of the knot. 
Alternatively, a jet may accelerate as it leaves the high density parent cloud and enters the lower density ambient medium \citep[see, e.g.][]{daG96}, leading to an underestimate of the age of the feature. 
With only two epochs, we cannot measure the velocity evolution of the jets, and therefore assume ballistic motion to calculate representative dynamical ages. 
Tables~2 and 3 list age estimates derived from the proper motions and positional offset from the driving source (or approximate source position in the case of HH~901 and HH~902). 
HH~666 is the longest outflow in our small sample, extending more than 3 pc in projection, and likely $\gtrsim 4$ pc in total length \citep[based on the estimated $\alpha \approx 30^{\circ}$; see also][]{smi04}. 
Several knots have derived ages $\ga 4000$ years, corresponding to $\sim 1$\% of the Class I lifetime derived for low-mass stars \citep[0.54 Myr,][]{eva09}. 
This lower limit on the fraction of the accretion age traced by the outflow does not account for the faster evolution of more massive protostars \citep[the best-fit model of the IR spectral energy distribution of the HH~666 driving source yields a protostar mass of $\sim 6.3$ M$_{\odot}$,][]{pov11,rei13}. It also does not include the larger age ($\sim45,000$ yr) of HH~666~D, which we estimate from the radial velocity and estimated tilt angle $\alpha\approx30^{\circ}$ as it falls outside the area imaged with WFC3. 
The driving source is detected at visible wavelengths \citep[see Figure~\ref{fig:hh666_pm} and][]{smi10}, indicating that it is no longer deeply embedded along our line of sight. 
Because HH~666 lives in the harsh radiative environment of the Carina nebula, it is unclear how much of the relatively low extinction to the driving source reflects protostellar evolution, clearing by the jet, or how much the envelope may have been eroded by external irradiation. 
Nevertheless, the high mass-loss rate of the inner jet (features HH~666~M and O) evident from bright H$\alpha$ and [Fe~{\sc ii}] emission demonstrates that intermediate-mass protostars can continue to accrete at a high rate, albeit sporadically, even as the protostellar envelope is dispersed. 

HH~901 and HH~902 have H$\alpha$ surface brightnesses similar to HH~666 and velocities a factor of $\sim 2$ lower. 
The dynamical ages for the most distant knots in HH~901 and HH~902 are a factor of $\sim 5$ younger than those for HH~666. 
Unlike HH~666, there is no direct detection of IR flux from the driving source of HH~901 or HH~902 \citep[see Section~\ref{s:results}, and][]{rei13}, leaving the mass and evolutionary stage of their driving sources poorly constrained. 
If these strong outflows are all driven by similarly evolved protostars, then the high mass-loss rates demonstrated by all three jets provides additional evidence that environmental differences strongly limit our detection sensitivity for the driving sources of HH~901 and HH~902.  

HH~1066 is most remarkable for its apparent youth. 
The dynamical ages of the bow shocks are $\sim 350$ and $\sim 550$ years for the eastern and western bow shocks, respectively, and the two newly discovered knots in the inner jet have dynamical ages of $\sim 35$ years (see Section~\ref{s:results}). 
Short wavelength emission from the driving source \citep[15.5 mag at J-band,][]{pov11} suggests that it is beyond the earliest evolutionary phase (Class 0), although evidence of an optically thick, nearly edge-on accretion disc \citep{rei13} suggests that it is still relatively young. 
Further monitoring of HH~1066 is required to assess whether it is entering a phase with a rapid succession of accretion bursts that will generate a continuous but clumpy stream of [Fe~{\sc ii}] emission from the source, similar to HH~666~M and O.

\subsection{Momentum injection}\label{ss:momentum} 

Jets may inject considerable momentum into the surrounding star forming environment. 
Adopting the average parameters for an outflow in Carina, 
$\dot{M_w} \sim 10^{-6}$ M$_{\odot}$ yr$^{-1}$ \citep{rei13} and 
$v \approx 150$ km s$^{-1}$, the average momentum flux from these HH jets is 
$\dot{p_w} = \dot{M_w} v \approx 1.5 \times 10^{-4}$ M$_{\odot}$ km s$^{-1}$ yr$^{-1}$.  
Over the course of the jet lifetime of $\sim 10^5$ yr, the total momentum injection for a single intermediate-mass protostar will be $p_w \approx 15$ M$_{\odot}$ km s$^{-1}$. 
This is similar to the momentum injection of $18$ M$_{\odot}$ km s$^{-1}$ that \citet{wal05} find for HH~46/47 which is driven by one of the two T Tauri stars in the 12 L$_{\odot}$ binary source \citep{rei00}.

Based on this rough estimate, it is natural to ask whether the momentum injection by jets from intermediate-mass protostars competes with the contribution from their more numerous low-mass counterparts. 
While the velocities we measure in the jets in Carina are similar to the velocities measured in jets driven by low-mass protostars \citep[$\sim 150$ km s$^{-1}$, see Figure~\ref{fig:vhist}, and][]{rei01}, the high densities found by \citet{rei13} suggest that their mass-loss rates are $100-1000$ times higher than those observed in low-mass stars \citep[e.g., the mass-loss rates calculated from the H$\alpha$ emission measure for the HH jets in Orion,][]{bal01}. 
Calculating the momentum flux as $\dot{M}v$, jets from intermediate-mass stars with similar velocities and higher densities therefore have a momentum flux that is $100-1000$ times higher for an individual jet. 

To estimate whether the total momentum injection from intermediate-mass protostars exceeds that of the more numerous low-mass sources, we must compare the relative numbers of low- and intermediate-mass stars with the relative strengths of the jets they drive. 
Assuming a Salpeter initial mass function \citep[IMF,][]{sal55}, 
\begin{equation}
N = \int_{M_1}^{M_2} \xi(M) dM = \int_{M_1}^{M_2} \xi_0 M^{-2.35} dM
\end{equation}
we find that $0.5-2$ M$_{\odot}$ stars outnumber $2-8$ M$_{\odot}$ stars by a factor of $\sim 6.5$, although the number will change depending on the IMF used \citep[e.g.][]{kro01}. 
If each individual intermediate-mass star injects 100 times the momentum of each low mass star, then the cumulative momentum injection rate of intermediate-mass stars is still $\sim 15$ times that of low-mass stars. 
In principle then, intermediate-mass stars may be an important source of momentum injection in massive star forming regions. 
However, there are many caveats to consider when interpreting the relative momentum injection of low- and intermediate-mass stars. 

This simple estimate does not account for the possibility that the jet mass-loss rate and velocity may be functions of both the protostellar mass and evolutionary stage. 
More massive protostars evolve faster than their low-mass counterparts, which means both that their outflows turn on earlier and persist for a shorter period of time. 
The early injection of momentum from intermediate-mass stars may be more influential to the star forming environment than the cumulative contribution of low-mass stars later in the evolution of the cloud as a whole. 
Variability and precession may also affect the momentum transfer efficiency 
as a recent burst from a precessing jet that impacts fresh gas may transfer more momentum to the environment than a more steady-state jet propagating into a cleared cavity.

All four of the fast, dense jets studied here have high momentum flux, despite apparent differences in the evolutionary stage of their driving sources (see Section~\ref{ss:ages}). 
Of the four outflows we observe, HH~666 has the largest dynamical age and lowest obscuration of the driving source \citep[which can be seen in the H$\alpha$ images presented by][]{smi10}, suggesting that it is the oldest, most evolved source in the sample. 
Despite this, HH~666 maintains the highest velocities of the four jets measured here, and one of the highest mass-loss rates of all of the jets in Carina \citep[see Table~4 in][]{smi10}. 
If the irradiated jets in Carina behave in the same way as their embedded counterparts, then these four jets contradict the \citet{bon96} and \citet{bel08} conclusion that more evolved sources drive less powerful outflows. 
However, the efficiency of momentum transfer to the environment complicates the comparison between the Carina jets and the CO outflows observed by e.g., \citet{bon96,bel08,cur10,van13}. 

\citet{bon96} measure the CO momentum flux near the driving source assuming that the combined momentum flux of the inner jet and ``classical'' CO outflow is conserved along the flow direction. 
We observe the protostellar jet directly in Carina, so it is critically important to understand the efficiency of momentum transfer between the jet and the ambient gas that is swept up in the CO outflow. 
\citet{che94} find that momentum transfer from jets with a Mach number $>6$ happens primarily at the bow shock. Even in the ionized gas of the H~{\sc ii} region where the sound speed, $c_{II} \approx 11$ km s$^{-1}$, the Mach number of these jets is $>10$. In the neutral environment of the dust pillar, the Mach number for the same outflow speed is even higher. 
This suggests that the majority of the momentum transfer to the environment will happen when the jet turns on, or in the case of episodic jets (see Section~\ref{ss:variability}), each time the jet turns up \citep[losing mass at a higher rate corresponding to an accretion burst, see][]{cro87}. 
Momentum transfer happening primarily at the bow shock is also consistent with the interpretation that the Carina jets can maintain relatively high velocities throughout because subsequent knots travel in the wake of previous ejecta (see Section~\ref{ss:structure}).

\subsection{Jet variability}\label{ss:variability}

Clumps in protostellar jets may result from variations in the initial ejection velocity or changes in the density of the ejected material. 
Velocity variations in the jet may create knotty structures that move along the flow axis with small (few 10s of km s$^{-1}$) velocity variations between them reflecting the variation in the ejection velocity \citep[e.g.][]{rag90,har93,daG94}. 
Alternately, a jet may show a clumpy morphology but have a smooth velocity structure that suggests changes in the amount of mass lost into the jet at a constant ejection velocity (as may be the case in HH~666~M, Reiter et al. in prep). 
As \citet{rei89} and \citet{har90} have pointed out, larger outbursts evident as bow shocks in protostellar jets may be a way to infer the time between FU Orionis outbursts. 
To first order, we can estimate the time between separate ejection events by taking the difference in the age of large neighboring clumps. 
Simply taking the age difference of subsequent knots, the time between ejection events in these jets appears to be on the order of $\sim 500-5000$ years. 
The average time between ejections for all four jets is $\sim 1500$ yr, on the order of the variability time-scale estimated for HH~46/47 \citet{rag92}, and the time between ejections found in HH jets driven by low-mass stars \citep[e.g.][]{rei89,har90,rei92}.

Time variable accretion and outflow will impact many of the parameters we have estimated from these jets. 
Clumpy structures in the jet suggest that the mass-loss rate in the jet is elevated for short periods of time, spending more time in a quiescent phase, losing mass at a lower rate. 
If we assume that the jet is ``on'' (losing mass at a high rate) for only a fraction of the jet lifetime of $\sim 10^5$ yr, then the total momentum injection from a single intermediate-mass protostar will be reduced compared to what we would estimate assuming the maximum mass-loss rate for the duration of the jet lifetime.

\section{Conclusions}\label{s:conclusions}

We present new proper motion measurements of four massive HH jets in the Carina nebula -- HH~666, HH~901, HH~902, and HH~1066. 
With a $\sim 4.4$ year baseline and assuming a distance of 2.3 kpc, we measure transverse velocities of $\sim 43-269$ km s$^{-1}$. 
Velocities for several knots fall among the speeds measured in jets from low-mass protostars \citep[$\sim 100-200$ km s$^{-1}$,][]{rei01}, somewhat slower than the high outflow velocities inferred from spectra by \citet{cor97} for Herbig Ae/Be stars. 
With these velocities \citep[which are within a factor of 2 of the velocity assumed by][]{smi10}, the estimated mass-loss rates of these jets are $\sim$ few $\times 10^{-6}$ M$_{\odot}$ yr$^{-1}$ \citep{rei13}, similar to the mass-loss rates observed in FU Orionis objects \citep{hc95}. 
The knots we use to measure proper motions in the HH jets in Carina lie far from the driving source, where processing through shocks has almost certainly reduced the outflow velocity. 

New spectra of HH~901, HH~902, and HH~1066, combined with previously published spectra of HH~666 \citep{smi04}, make it possible to measure the tilt angles and 3D space motions of these jets. HH~901 and HH~902 both lie close to plane of the sky with tilt angles $\alpha \sim 20^{\circ}$. 
HH~1066 is tilted $\sim 20-45^{\circ}$ relative to the plane of the sky, with the redshifted western lobe of HH~1066 pointing toward, and possibly colliding with the neighboring pillar. 
Large proper motions measured for HH~666 suggest a tilt angle away from the plane of the sky of $\sim 35^{\circ}$. 

HH~1066 is the only jet in the sample that shows evidence of a recent change in the mass-loss rate. 
A difference image of the two epochs reveals two new blobs along the jet axis (and the high-velocity eastern bow shock). 
These knots are confused with the bright edge of the pillar, but an offset of $\sim 0.1$\arcsec\ in an intensity tracing suggests they were ejected in the last $\sim 35$ years (assuming a velocity of 250 km s$^{-1}$). 
Additional epochs are required to measure the proper motions of these new blobs, but they nevertheless provide strong evidence for time-variable accretion and outflow. 

The high densities inferred by \citet{rei13} together with the velocities we present here suggest that intermediate-mass stars inject substantial momentum into their natal star forming regions, contributing as much if not more momentum than their more numerous low-mass counterparts. 
These irradiated jets also add to evidence that momentum transfer to the environment is episodic and dominated by the initial collision with the ambient medium \citep{nar96,arc01}. 

Measuring velocities provides one of the crucial constraints for measuring the time-variable mass-loss history of these jets.
Evidence for a recent change in the mass-loss rate in HH~1066 points to the variable nature of mass accretion, suggesting that these sources provide a unique way to constrain the duty cycle and contrast in mass-loss-rate between the quiescent and outbursting states.


\section*{Acknowledgments}
We thank the anonymous referee for comments that improved the quality of the paper. 
We also wish to thank Pat Hartigan for helpful conversations and allowing us to consult his code to check against our proper motion measurement algorithm, and Rob Simcoe for assistance with FIRE data processing. 
MR would like to thank Tim Axelrod, Adam Burgasser, and Brant Robertson for helpful discussions. 
Support for this work was provided by NASA grants AR-12155 and GO-13391 from the Space Telescope Science Institute. 
This work is based on observations made with the NASA/ESA Hubble Space Telescope, obtained from the Data Archive at the Space Telescope Science Institute, which is operated by the Association of Universities for Research in Astronomy, Inc., under NASA contract NAS 5-26555. These observations are associated with programs GO/DD~12050 and GO~10241 and 10475.
We acknowledge the observing team for the HH~901 mosaic image:
Mutchler, M., Livio, M., Noll, K., Levay, Z., Frattare, L., Januszewski, W., Christian, C., Borders, T.


\label{lastpage}


\end{document}